\documentclass[preprint,authoryear,12pt]{elsarticle}

\usepackage{amssymb}
\usepackage{color}
\usepackage{amssymb}
\usepackage{bbm}
\usepackage{amsthm}
\usepackage{amsmath}
\usepackage{epsfig}
\usepackage{ae,aecompl}
\usepackage[version=3]{mhchem}
\usepackage{appendix}
\usepackage{setspace}
\usepackage{epstopdf}
\newcommand{\indic}[1]{1\hspace{-2.1mm}{1}_{\{#1\}}}

\newtheorem{thm}{Theorem}
\newtheorem{lem}{Lemma}
\newtheorem{prop}{Proposition}
\newtheorem{cor}{Corollary}
\theoremstyle{definition}

\newtheorem{rem}{Remark}

\begin{document}

\begin{frontmatter}



\title{An Optimal Multiple Stopping Approach to Infrastructure Investment Decisions}


\author[eric]{Eric Dahlgren}
\ead{ed2405@columbia.edu}
\address[eric]{Lenfest Center for Sustainable Energy, Department of Earth \& Environmental Engineering, Columbia University, New York, NY 10027}
\author[tim]{Tim Leung\corref{cor1}}
\ead{leung@ieor.columbia.edu}
\address[tim]{Department of Industrial Engineering \& Operations Research, Columbia University, New York, NY 10027}
\cortext[cor1]{Corresponding author}

\begin{abstract}
The energy and material processing industries  are traditionally characterized by very large-scale physical capital that is custom-built with long lead times and long lifetimes. However, recent technological advancement in low-cost automation has made possible the parallel operation of large numbers of small-scale and modular production units. Amenable to mass-production, these units can be more rapidly deployed but they are also likely to have a much quicker turnover. Such a paradigm shift motivates the analysis of the combined effect of lead time and lifetime on infrastructure investment decisions.  In order to value  the underlying real option, we introduce an optimal multiple stopping approach that accounts for operational flexibility, delay induced by lead time, and multiple (finite/infinite) future  investment opportunities.  We provide an analytical characterization of the firm's value function and optimal stopping rule. This leads  us to develop an iterative numerical scheme,  and examine how the investment decisions depend  on  lead time and lifetime, as well as other parameters. Furthermore, our model can be used to  analyze the critical investment cost that makes small-scale (short lead time, short lifetime) alternatives  competitive with traditional large-scale infrastructure.
\end{abstract}

\begin{keyword}
optimal multiple stopping \sep real option \sep  infrastructure investments \sep lead time \sep operational flexibility

\MSC[2010] 60G40 \sep 62L15\sep 62L20
\JEL C41 \sep G13 \sep H54

\end{keyword}

\end{frontmatter}

\section{Introduction}
The energy and material  (e.g. water and petrochemicals) processing industries are traditionally characterized by very large unit-scale physical capital. For example, individual electric power generators rated in the 100s of MW, distillation towers in refineries measuring 100,000s of barrels-per-day of capacity, and mining trucks capable of hauling 400 tons of ore  are common sizes in their respective industries. However, the recent emergence of low-cost automation technologies makes a modular, small-scale, and potentially distributed  approach possible with  comparable aggregate production capacity.  This calls for a re-examination of the current ``bigger-is-better'' paradigm. Indeed, abandoning \emph{economies of unit scale} in favor of \emph{economies of mass-production} presents several opportunities, as discussed in  \cite{Dahlgren}.   In contrast to the very long lifetimes of 25 years or more for typical large-scale capital, small and mass-produced equipment might be endowed with a much shorter physical lifespan. This could be either by design in construction or in operation with a limited maintenance schedule. Furthermore, with a smaller unit scale comes the ability to build to stock and drastically shorten lead times between the investment and operation. These factors provide the firm with additional flexibility to  engage and disengage a given activity, which should be accounted for in the investment valuation and decision.

Motivated by this anticipated paradigm shift, we introduce here a framework that incorporates operational flexibility, lead time, capital lifespan, and multiple (finite/infinite) future  investment opportunities. To this end, we formulate an  optimal multiple stopping problem, where the firm maximizes the expected discounted  reward from sequential investments.  In particular, the project's reward function captures the operational flexibility of temporarily suspending production to avoid negative cash flows. Additionally, it depends explicitly on two crucial elements, namely,  lifetime and  lead time. Under capacity constraint, the  firm's consecutive investments are separated (or refracted) by the capital lifetime.  Hence, the firm's investment decision bears similarity to the valuation of a forward-starting swing call option written on the reward.

 Our main  result is the characterization of the firm's value function and optimal stopping rule.  The optimal timing  for multiple  investments is  described by a  sequence of critical price thresholds. Furthermore, these thresholds are shown to be  decreasing and converge to that corresponding to the case with infinite investment opportunities. Our analysis lends itself to an iterative algorithm that  numerically solves for the optimal value function and all exercise thresholds, as opposed to the simulation approach commonly found in  existing literature for swing-type options \citep{Meinshausen,Chiara,Bender}. We examine the impacts of  lead time and lifetime, as well as other parameters,  on the  investment decisions. Moreover, our model is also useful for   analyzing the critical investment cost that makes small-scale (short lead time, short lifetime) alternatives  competitive with traditional large-scale infrastructure.

As is well known,  the real option approach to investment decisions can increase the project  value above and beyond those rooted in classical net present value (NPV) arguments  \citep{Sick_Gamba,DixitPindyck}. In its inherent myopia, the  NPV approach falls short of capturing the flexibility of timing since it, at best, only gives the decision maker an indication of whether or not  a single investment should be made at the current time. Standard real investment option analysis addresses this limitation. Examples of the implementation of such an analysis regarding individual investment decisions can be found in, among others,  \cite{Kaslow199460,Frayer200140,Lumley01112001,Carelli2010403} and \cite{Westner201231}. Nevertheless, the valuation of multiple consecutive  investments has not received a great deal of attention in the industries in question. In  the current paradigm, the individual investment is  typically both large and long-lived, so  the firm's  next investment decision, including capacity replacement,  will be an issue of the far future. With such a long horizon, discounting would reduce  future cash flows to  minimal  present values. However, with shorter lead times and  lifetimes, future investments can potentially  have significant bearing  on current capital budgeting decisions, as we will show in this paper.

{Our valuation approach accounts for multiple   and possibly infinite  future investment options, which in turn allows for a comparison of  capital investments  with different lifetimes. To illustrate, let us compare two projects with different lifetimes, namely, one with  $10$ years and another with $3$ years. To address the difference in investment horizon, one approach would be to consider a longer horizon of 30 years, the smallest common multiple of 3 and 10, and compare   the corresponding net present values. However, this method implicitly assumes seamless consecutive investments at the end of each lifetime. This may not be optimal to the firm since it ignores the embedded timing option in each investment decision.}

Another  feature of our valuation framework is to incorporate the flexibility to delay future investments depending on market conditions. Another advantage over the pairwise comparison is that we can easily examine how sensitive the investment timing is with respect to  different model parameters, especially for all lead times and lifetimes. We do acknowledge that durability of capital is a path-dependent variable, leading to a variable lifetime. In contrast, our approach  uses a fixed lifetime independent of the frequency of operation, and it may  favor  the posited status quo of capital with a longer prescribed lifetime. Our choice of a fixed lifetime, albeit a simplification, results from the trade-off between model tractability  and realistic details.

The use of swing options has been most prevalent in energy delivery contracts   where the holder has the right to alter, or `swing',  volumes up or down at the start of each time period \citep{Jaillet01072004,Deng2006940}.  Rather than fixed-period contracts, the only constraint on the timing in our setting is a minimum refraction time between consecutive investments. As such, the valuation of the swing contract can be formulated as a refracted optimal multiple stopping problem (see e.g. \cite{MAFI:MAFI331,carmona2008optimal}). Other financial applications  involving multiple exercises include  employee stock option valuation \citep{LeungSircarESO_MF09,GrasselliHenderson2008}, and the operation of a physical asset \citep{Ludkovski}.

This paper is structured as follows. In Section \ref{sec:problem_formulation} we introduce the investment decision as a general optimal multiple stopping problem. We also introduce a reward function that incorporates the flexibility to temporarily shut down production to avoid a negative cash flow. In the following section we state and prove sufficient conditions on a general reward function for the multiple stopping problem to have a well defined stopping boundary. We also present an algorithmic approach to finding the solution. In Section \ref{sec:application_to_infrastructure} we further motivate and analyze the specific reward function in the context of basic infrastructure investments and present numerical results. These results include a comparison of investment scenarios with short lifetimes and short lead times versus a traditional scenario with comparatively long lifetimes and lead times. Finally, concluding remarks and potential extensions are discussed in Section \ref{extensions}.

\section{Problem Formulation}
\label{sec:problem_formulation}
We consider a firm that has the ability to invest in capital equipment that produces a single good in a given market. Furthermore, we assume that this firm  is acting as a price taker in this market, with a finite capacity constraint. These conditions imply that the investment decision will be of `bang-bang' type, that is, either invest up to this limit or not at all.  From an operational cost perspective, we are agnostic to whether the total capacity comes in one big unit or in 10,000 smaller units. Under such circumstances, we can without loss of generality analyze the investment timing on a per-unit-capacity basis.

We consider an exogenous and deterministic investment cost, $I(T,\nu)$, which depends on lifetime, $T$, and lead time, $\nu$, among other factors. These two parameters, both considered deterministic, also affect directly the net present value, $\psi$, of a single investment. Moreover, we consider the discount rate, $r$, used by the firm to also be exogenous and constant. In the background, we fix a complete probability space $(\Omega, \cal{F}, \mathbb{P})$,  equipped with a filtration $\mathbb{F}=({\cal{F}}_t)_{t\geq 0}$. Let  $(X_t)_{t\ge 0}$  be an $\mathcal F_t$-adapted output price process. An investment generates a random cash flow-process of the form $(f\left(X_t\right))_{t\ge 0}$, where $f$ is a function known to the firm.

The expected discounted stream of future cash flows, minus the initial investment cost, yields the  net present value
\begin{equation}
\label{net_present_value}
\psi(x;T,\nu) = -I(T,\nu) +\int_{\nu}^{\nu+T}e^{-r t}\,\mathbb{E}\left\{f\left(X_t^{0,x}\right)\right\}\,dt,\qquad 0<x<\infty,
\end{equation}
where we have introduced the conditional notation $X_t^{0,x}\equiv \left\{X_t|X_0=x\right\}$. The expression in (\ref{net_present_value}) helps clarify the role of the lead time $\nu$ as the time between the expenditure $I$ and start of operation and access to the cash flow $f\left(X_t\right)$.

The limitation on the amount of capacity that can be active at any given point in time, as implied by the assumption of the role as a price taker discussed above, naturally introduces the notion of a refraction time. That is, two consecutive investments have to be separated with at least a time $T$, the lifetime of the capital. With this being the only restriction on the investment strategy of the firm, the value $v^{(k)}(x;T,\nu)$ of $k$ consecutive investments can be formulated as an optimal multiple stopping problem
\begin{equation}
\label{multiple_stopping_problem}
v^{(k)}(x;T,\nu) = \sup_{\substack{\vec{\tau}\in {\cal{S}}^{k}}}\mathbb{E}\left\{\sum_{i=1}^ke^{-r \tau_i}\psi(X_{\tau_i}^{0,x},T,\nu)\right\},\qquad 0<x<\infty.
\end{equation}
The set ${\cal{S}}^{k}$ is the set of all refracted stopping times
$\vec{\tau}=(\tau_1,\tau_2,\dots,\tau_k)$, i.e. $\tau_i-\tau_{i-1}\geq T$ for $i=2,3\dots,k$.  Under this requirement, nothing prevents the firm from having two consecutive investments operating with a seamless transition. This is because the decision to invest in future capital can be made before the current investment expires. Provided that  a least upper bound in (\ref{multiple_stopping_problem}) actually exists, we can include in ${\cal{S}}^{k}$ the stopping vector where $\tau_{p}=\tau_{p+1}=\dots=\tau_k=\infty$, $p\leq k$. For such a stopping rule,  $v^{(k)}$ can be interpreted as the value of the contract when not every exercise is being called.

The tacit assumption that the supremum in (\ref{multiple_stopping_problem}) exists implies that $e^{-rt}\psi(X_t;T,\nu)$ is integrable and $\lim_{t\to\infty}\mathbb{E}\left\{e^{-rt}\psi(X_t;T,\nu)\right\}=0$ for every choice of $T$ and $\nu$. For brevity we will suppress the parameters $T$ and $\nu$ in $\psi$, $v^{(k)}$ and $I$, unless such a dependence is specifically required. Since the firm is free to choose the timing of every investment, up to the condition of a refraction time $T$, the following proposition highlights the optionality embedded in the investment decision. A proof is provided in \ref{app:proof_prop}. 

\begin{prop}
\label{prop:positive_reward}
The value function $v^{(k)}(x)$, $k\geq 1$, satisfies,
\begin{equation}
v^{(k)}(x) =\sup_{\substack{\vec{\tau}\in {\cal{S}}^{k}}}\mathbb{E}\left\{\sum_{i=1}^ke^{-r \tau_i}\psi^+(X_{\tau_i}^{0,x})\right\},\label{eqnprop1}
\end{equation}
with $\psi^+(x) = \max\{0,\psi(x)\}$.
\end{prop}

Proposition \ref{prop:positive_reward} reveals that it is never optimal to make the investment in the negative region of $\psi(x)$.  This is because the firm has the flexibility to delay investment, and therefore, can always avoid value destruction.

We now consider  the formulation in (\ref{net_present_value}) and (\ref{multiple_stopping_problem}) under the geometric Brownian motion (GBM) model. Specifically, we model an output price process by the  stochastic differential equation (SDE):
\begin{equation}
\label{gbm}
dX_t = \alpha X_t\, dt + \sigma X_t\, dB_t, \qquad X_0 \in (0, \infty),
\end{equation}
where $(B_t)_{t\ge 0}$ is a standard Brownian motion. The drift rate $\alpha$ and the variance parameter $\sigma$ are both assumed constant. We denote $\mathbb{F}=({\cal{F}}_t)_{t\geq 0}$ to be  the filtration generated by $B$.

From the firm's perspective, we assume that the investment cash flow  is  the difference between an uncertain output price $X_t$, modeled by (\ref{gbm}), and a constant operational cost $c$. (In Section \ref{sec:application_to_infrastructure} we show that such a formulation is equivalent to a cash flow being the difference of two GBMs, possibly correlated, through a change of measure.) Moreover, the firm has the \emph{operational flexibility} to  temporarily suspend production if cost exceeds output price. Hence, the effective investment cash flow $f(X_t^{0,x})$ at any future time $t$ is given by
\begin{equation}
\label{cash_flow}
f(X_t^{0,x})= \left(X^{0,x}_t-c\right)^+,
\end{equation}
where $\left(X_t-c\right)^+ = \max\{0,X_t-c\}$. As such, the expectation of  $f(X_t^{0,x})$ in (\ref{cash_flow}) can be seen as the price of a European call option on $X$ with a strike price $c$ and maturity $t$ (see \cite{McDonald1985}), namely,
\begin{equation*}
\mathbb{E}\left\{\left(X_t^{0,x}-c\right)^+\right\}= x\Phi(d_+(t))e^{\alpha t}-c\Phi(d_-(t)),
\end{equation*}
where $\Phi$ is the standard  normal cumulative distribution function, and where
\begin{equation*}
d_{\pm} (t)= \left[\ln\left(\frac{x}{c}\right)+\left(\alpha\pm\frac{1}{2}\sigma^2\right)t\right]/
\sigma\sqrt{t}\,.
\end{equation*}
Applying this to \eqref{net_present_value}, the reward function $\psi(x)$  becomes
\begin{equation*}
\psi(x) = -I + \int_{\nu}^{\nu+T}\left(x\Phi(d_+(t))e^{\alpha t}-c\Phi(d_-(t))\right)e^{-rt}dt\,.
\end{equation*}

Under this setting,  we  recast the optimal multiple stopping problem as a sequence of optimal single stopping problems. More precisely, we express the value $v^{(k)}(x)$ as
\begin{equation}
\label{single_stopping}
v^{(k)}(x) = \sup_{\substack{\tau_k\in {\cal{S}}}}\mathbb{E}\left\{e^{-r \tau_k}\psi^{(k)}(X_{\tau_k}^{0,x})\right\}.
\end{equation}
The set ${\cal{S}}$ is the set of all $\mathbb{F}$-stopping times and
\begin{equation*}
\psi^{(k)}(x) = \psi(x)+e^{-r T}\mathbb{E}\left\{v^{(k-1)}(X_{T}^{0,x})\right\},\ \ v^{(0)}\equiv 0\,.
\end{equation*}
 By construction, the admissible stopping times $(\tau_1, \ldots, \tau_k)$ are again refracted  by at least $T$ (years). By standard optimal stopping theory, the Snell envelope $\left(e^{-r t}v^{(k)}(X_t)\right)_{t\geq 0}$ is constructed to be the smallest supermartingale dominating the process $\left(e^{-r t}\psi^{(k)}(X_t)\right)_{t\geq 0}$, for every $k$. For further  details regarding this approach under a more general framework, we refer to \cite{MAFI:MAFI331}.

\section{Analytical Results}
\label{sec:solution_approach}
In this section we provide an analytical study  for the optimal stopping problem in (\ref{single_stopping}).  Our main result is the characterization of the value function and optimal stopping rule,  for every $k\geq 1$  opportunities to invest (see Theorem \ref{main_theorem}). This leads us to develop an iterative approach to finding the value functions $v^{(k)}(x)$ and the optimal exercise boundaries $x^*_k$ (see  Corollary \ref{algorithm_prop}), which will be the foundation for our numerical algorithm to be presented in Section \ref{sec:implementation}.

In the context of real investments, let  us  discuss some conditions on the general  reward function $\psi$. First, we assume that $\psi$ is continuous, increasing and sufficiently smooth. Additionally, we assume that there is a unique break-even point, $x_0>0$, where $\psi(x_0)=0$, and $\psi(x)< 0$ for $x< x_0$ and $\psi(x)> 0$ for $x>x_0$. Finally, we note that $\psi(x)$ is bounded by some affine function of $x$. Such a bound, together with an assumption that the discount rate $r$ exceeds the drift rate $\alpha$ of the underlying process, ensures that perpetual waiting will lead to zero expected reward, i.e. $\lim_{t\to\infty}\mathbb{E}\left\{e^{-r t}\psi(X_t)\right\}=0$. Under these conditions,  we will first analyze the real option  problem with a single investment opportunity, which will be the building block for solving the  optimal multiple stopping problem.

\subsection{Optimal Single Stopping Problem}
\label{sec:opt_stopping_rule}We first consider the optimal single stopping problem
\begin{equation}
\label{single_stopping_temp}
v^{(1)}(x) = \sup_{\substack{\tau_1\in {\cal{S}}}}\mathbb{E}\left\{e^{-r \tau_1}\psi(X_{\tau_1}^{0,x})\right\}.
\end{equation}
This problem is similar to the pricing of a perpetual American   option with  the payoff  $\psi$. Consider a candidate  stopping time for  problem (\ref{single_stopping_temp}):
\[\tau_{x_1^*} = \inf\{t\geq 0\,|\,X_t^{0,x}\geq x_1^*\,,\,x>0\,\},\] with threshold $x_1^*>0$.    By the well-known Laplace transform of the first passage time of $X$, see e.g. \citep[p.346]{Shreve}, we have for  $x\leq x_1^*$,
\begin{equation}
\label{Laplace}
\mathbb{E}\left\{e^{-r \tau_{x_1^*}}\psi(X^{0,x}_{\tau_{x_1^*}})\right\}=
\psi(x_1^*)\left(\frac{x}{x_1^*}\right)^{\gamma},
\end{equation}where   

\begin{equation}\notag
\gamma = \frac{1}{2} - \frac{\alpha}{\sigma^2}+ \sqrt{\left(\frac{1}{2} - \frac{\alpha}{\sigma^2}\right)^2+\frac{2r}{\sigma^2}}\,.
\end{equation}
Note that  the condition $r>\alpha$ implies that $\gamma>1$.
If  $x\geq x_1^*$, then $\tau_{x_1^*}=0$ and $v^{(1)}(x)=\psi(x)$. From (\ref{Laplace}) we see that a necessary condition for $x_1^*$ to be an optimal stopping boundary is that $x_1^*$ maximizes $\psi(x)/x^{\gamma}$. The linear bound on $\psi(x)$ ensures the existence of such a maximum.

The first order condition for a maximum at $x=x_1^*$ is given by
\begin{equation}
\left.\frac{d}{dx}\frac{\psi(x)}{x^{\gamma}}\right|_{x=x^*_1} = \frac{x^*_1\psi '(x^*_1)-\gamma\psi(x^*_1)}{(x^*_1)^{\gamma+1}} \equiv -\frac{\Lambda\psi(x^*_1)}{(x^*_1)^{\gamma+1}} = 0\ \ \ \Leftrightarrow\ \ \ \Lambda\psi(x^*_1) = 0, \label{Lambda}
\end{equation}
where we have introduced the operator notation $\Lambda = \left(\gamma - x\frac{d}{dx}\right)$. The second order condition, sufficient to prove a maximum together with the first order condition above, can then be stated as
\begin{equation*}
\left.\frac{d^2}{dx^2}\frac{\psi(x)}{x^{\gamma}}\right|_{x=x^*_1} = -\frac{\frac{d}{dx}\Lambda\psi(x^*_1)}{(x^*_1)^{\gamma+1}} < 0\ \ \Leftrightarrow\ \ \frac{d}{dx}\Lambda\psi(x^*_1)>0.
\end{equation*}
Note that a maximum of $\psi(x)/x^{\gamma}$ at $x_1^*$ is not a sufficient condition for an optimal stopping rule $\tau_{x_1^*}$ for a general reward function $\psi$. One would also have to prove that $\left(e^{-r t}v^{(1)}(X_t)\right)_{t\geq 0}$ satisfies the supermartingale property for this choice of $\tau_{x_1^*}$. The following lemma gives the sufficient conditions for $\tau_{x_1^*}$ to be an optimal stopping rule for problem  (\ref{single_stopping_temp}).

\begin{lem}
\label{Lemma_1}
Let $\psi:\mathbb{R}^+\to\mathbb{R}$ be a reward function in the single stopping problem (\ref{single_stopping_temp}). If $x_1^*$ is a global maximum for $\psi(x)/x^{\gamma}$ on $\mathbb{R}^+$ and if
\begin{equation}
\label{cond_lemma}
\frac{d}{dx}\Lambda \psi(x)\equiv (\gamma-1)\psi '(x)-x\psi ''(x)\geq 0,\quad x\geq x_1^*,
\end{equation}
then
\begin{equation*}
v^{(1)}(x) = \psi(x\vee x_1^*)\left[1\wedge \left(\frac{x}{x_1^*}\right)^{\gamma}\right],\ \ \ x\in\mathbb{R}^+,
\end{equation*}
where $v^{(1)}(x)$ is continuous on $\mathbb{R}^+$.
\end{lem}

\begin{rem}
The first order condition $\Lambda\psi(x_1^*)=0$ in (\ref{Lambda}), together with the condition that $(d/dx)\Lambda \psi(x)\geq 0$,  $x\geq x_1^*$, bounds the second derivative (convexity) of the reward function $\psi(x)$ for large $x$. Up to the condition of a maximum of $\psi(x)/x^{\gamma}$ at $x=x_1^*$, the behavior of the function $\psi(x)$ to the left of $x=x_1^*$ is irrelevant. The conditions in Lemma \ref{Lemma_1} are therefore less restrictive than imposing that the drift term of $e^{-r t}\psi(X_t)$ be monotone for all $X_t$, as presented in \citep[pp.128-130]{DixitPindyck} as a part of the sufficient conditions for a connected stopping boundary at $x=x_1^*$ for a perpetual American call on $\psi(x)$.
\end{rem}

\noindent {\bf Proof:} From (\ref{Laplace}) we know that $\hat{v}(x)$, where
\begin{equation}
\label{solution_subset}
\hat{v}(x) =
\begin{cases}
\psi(x_1^*)\left(\frac{x}{x_1^*}\right)^{\gamma}, & x< x_1^*,\\
\psi(x),& x\geq x_1^*,
\end{cases}
\end{equation}
is a candidate for the solution. From the conditions  on $\psi(x)$, it follows that, for  $x\geq x_1^*$,
\begin{eqnarray}
\label{supermartingale_1}
\Lambda\psi(x) &=& \gamma\psi(x) - x\psi '(x)\geq 0,\\
\label{supermartingale_2}
\frac{d}{dx}\Lambda\psi(x) &=& (\gamma-1)\psi '(x) - x\psi ''(x)\geq 0.
\end{eqnarray}
Looking at  the drift term\footnote{For a  general underlying process $X$, a localization procedure may be applied to eliminate the martingale term of $de^{-r t}\hat{v}(X_t)$ upon expectation.} of $e^{-r t}\hat{v}(X_t)$,
\begin{eqnarray}
\mathbb{E}\left\{de^{-r t}\hat{v}(X_t)\right\} &=& e^{-r t}\frac{\psi(x_1^*)}{(x_1^*)^{\gamma}}\left[-r +\alpha \gamma +\frac{1}{2}\sigma^2\gamma(\gamma-1)\right] \indic{X_t\leq x_1^*}dt\nonumber\\
\label{supermartingale_3}
&&+e^{-r t}\left[-r \psi(X_t)+\alpha X_t \psi '(X_t)+\frac{1}{2}\sigma^2X_t^2\psi ''(X_t)\right] \indic{X_t\geq x_1^*}dt,
\end{eqnarray}
 we observe that the first term vanishes identically, and  the second term is non-positive following from  (\ref{supermartingale_1}) and (\ref{supermartingale_2}).  Hence, $\left(e^{-r t}\hat{v}(X_t)\right)_{t\geq 0}$ is a supermartingale, which implies  that
\begin{equation}
\hat{v}(x) =\mathbb{E}\left\{ e^{-r (0\wedge \tau)}\hat{v}(X^{0,x}_{0\wedge \tau})\right\} \geq \mathbb{E}\left\{e^{-r (t\wedge \tau)}\hat{v}(X^{0,x}_{t\wedge \tau})\right\},\ \ \tau\in{\cal{S}}.\label{proofaa}
\end{equation}

The linear bound on $\psi(x)$ implies that $e^{-r (t\wedge \tau)}\hat{v}(X^{0,x}_{t\wedge \tau})$ is integrable. Also, taking the limit $t\to \infty$ in \eqref{proofaa} and maximizing over $\tau$ yields:
\begin{equation}
\label{greater_than}
\hat{v}(x)\geq \sup_{\substack{\tau\in {\cal{S}}}}\mathbb{E}\left\{e^{-r \tau}\psi(X^{0,x}_{\tau})\right\}.
\end{equation}
Conversely, choosing the specific stopping time $\tau=\tau_{x_1^*}$, the process
$\left(e^{-r (t\wedge \tau_{x_1^*})}\hat{v}(X_{t\wedge \tau_{x_1^*}})\right)_{t\geq 0}$ is a martingale by construction and therefore
\begin{align}
\hat{v}(x) &=\mathbb{E}\left\{e^{-r (t\wedge \tau_{x_1^*})}\hat{v}(X_{t\wedge \tau_{x_1^*}})\right\}= \mathbb{E}\left\{e^{-r \tau_{x_1^*}}\hat{v}(x^*)\right\}\notag\\
&=\mathbb{E}\left\{e^{-r \tau_{x_1^*}}\psi(x^*)\right\}\leq \sup_{\substack{\tau\in {\cal{S}}}}\mathbb{E}\left\{e^{-r \tau}\psi(X^{0,x}_{\tau})\right\}.\label{less_than}
\end{align}
The expressions in (\ref{greater_than}) and (\ref{less_than}) together give the desired result,
\begin{equation*}
\hat{v}(x) = v(x)= \sup_{\substack{\tau\in {\cal{S}}}}\mathbb{E}\left\{e^{-r \tau}\psi(X^{0,x}_{\tau})\right\}. \ \ \ \square
\end{equation*}

\subsection{Optimal Multiple Stopping Problem}
With Lemma \ref{Lemma_1} and Proposition \ref{prop:positive_reward} we can now state the main result. See \ref{app:proof_thm} for a proof.

\begin{thm}
\label{main_theorem}
Let $\psi:\mathbb{R}^+\to\mathbb{R}$ be a reward function with a break-even point  $x_0$. If $\Lambda \psi (x)$ is convex for $x\in(x_0,\infty)$, with $\Lambda \psi (x)$ increasing for large $x$, then, for every $k\geq 1$, there exists an $x^*_k>x_0$ such that
\begin{equation}
\label{main_result}
v^{(k)}(x) = \psi^{(k)}(x\vee x^*_k)\left[1\wedge \left(\frac{x}{x_k^*}\right)^{\gamma}\right],\ \ k\geq 1,
\end{equation}
where
\begin{equation}
\label{intermediate_reward}
\psi^{(k)}(x) = \psi(x)+e^{-r T}\mathbb{E}\left\{v^{(k-1)}(X^{0,x}_{T})\right\}.
\end{equation}
Moreover, the sequence $\left(x^*_k\right)_{k\geq 1}$ is strictly decreasing, and $\left(v^{(k)}\right)_{k\geq 1}$ is a strictly increasing sequence of continuous functions on $\mathbb{R}^+$. Also, for any bounded subset $D\subset \mathbb{R}^+$ there exists a constant $K_D$, such that $v^{(k)}(x)\leq K_D$, for $x\in D$ and $k\geq 1$.
\end{thm}

From a computational perspective it is convenient to introduce the auxiliary function $u^{(k)}$ through
\begin{equation}
\label{u_def}
u^{(k)}(x) = \Lambda v^{(k)}(x) = \Lambda\psi^{(k)}(x)\indic{x\geq x_k^*}.
\end{equation}
With the conditions imposed on $\psi(x)$ (bounded by a linear function, convexity of $\Lambda\psi(x)$ on $(x_0,\infty)$ and $\Lambda\psi(x)$ being increasing for large $x$) one can, in a similar way as for $v^{(k)}$ in the proof of Theorem \ref{main_theorem}, show that $\left(u^{(k)}\right)_{k\geq 1}$ is an increasing sequence of continuous functions bounded on every bounded subset $D\subset\mathbb{R}^+$. Given the function $u^{(k)}(x)$ we can reconstruct the value function $v^{(k)}(x)$ through
\begin{equation}
\label{value_defined_through_u}
v^{(k)}(x) = x^{\gamma}\left(\frac{\psi^{(k)}(x_k^*)}{(x_k^*)^{\gamma}} - \int_0^xy^{-\gamma-1}u^{(k)}(y)dy\right).
\end{equation}
Note that since $u^{(k)}(x)=0$ for $x<x^*_k$ there are no convergence issues in (\ref{value_defined_through_u}). As a consequence of Theorem \ref{main_theorem} we have the following result.

\begin{cor}
\label{algorithm_prop}
The functions $u^{(k)}(x)$, for $k=1,2,\dots$, satisfy
\begin{equation}
\label{u_calc}
u^{(k)}(x) = \left(\Lambda\psi(x)+e^{-r T}\mathbb{E}\left\{u^{(k-1)}(X_T^{0,x})\right\}\right)\indic{x\geq x_k^*},\ \ u^{(0)}(x)\equiv 0,
\end{equation}
The boundary point $x_k^*$ is the unique solution to
\begin{equation}
\label{boundary_point}
\begin{cases}
\Lambda\psi(x_k^*)+e^{-r T}\mathbb{E}\left\{u^{(k-1)}(X_T^{0,x_k^*})\right\} & =0,\\
\left.\frac{d}{dx}\left(\Lambda\psi(x)+e^{-r T}\mathbb{E}\left\{u^{(k-1)}(X_T^{0,x})\right\} \right)\right|_{x = x_1^*}&>0.
\end{cases}
\ \ \ \square
\end{equation}
\end{cor}
Corollary \ref{algorithm_prop}, together with equation (\ref{value_defined_through_u}), outlines the inductive algorithm used to find the solution to (\ref{single_stopping}). Details of the implementation are found in the next section.

It now remains to consider the behavior of the solution to the optimal multiple stopping problem in the limit of infinitely many exercise rights. That is, we will investigate the value function
\begin{equation}
\label{infinite_stopping_problem}
v^{(\infty)}(x) = \sup_{\substack{\vec{\tau}\in {\cal{S}}^{\infty}}}\mathbb{E}\left\{\sum_{n\geq 1}e^{-r \tau_n}\psi(X_{\tau_n}^{0,x})\right\}.
\end{equation}
Since $\left(e^{-r t}aX_t\right)_{t\geq 0}$, where $ax>ax-b\geq \psi(x)$, is a supermartingale for all $X_t$ and since the refracted stopping times $(\tau_n)_{n\geq 1}$ satisfy $\tau_n\geq (n-1)T$, we have
\begin{equation}
\label{v_bound}
v^{(\infty)}(x)\leq \mathbb{E}\left\{\sum_{n=1}^{\infty}e^{-r \tau_n}aX_{\tau_n}^{0,x}\right\} \leq \sum_{i=0}^{\infty}\mathbb{E}\left\{e^{-r iT}aX_{iT}^{0,x}\right\} = \frac{ax}{1-e^{-(r-\alpha)T}}.
\end{equation}
That is, $v^{(\infty)}(x)$ is bounded on every bounded subset of $\mathbb{R}^+$. Defined through an integral in (\ref{infinite_stopping_problem}), this bound ensures continuity of $v^{(\infty)}(x)$ on every bounded subset of $\mathbb{R}^+$. Finally, we state the following convergence result.

\begin{prop}
The sequence $v^{(k)}(x)$ converges uniformly to $v^{(\infty)}(x)$ on every bounded interval of $\mathbb{R}^+$.
\end{prop}
\noindent {\bf Proof:} Without loss of substantial generality, we assume that $x\le M$. Let $(\tau^*_{n,\infty})_{n\geq 1}$ be an optimal stopping rule for the value function $v^{(\infty)}(x)$ in (\ref{infinite_stopping_problem}). Employing a similar argument as in (\ref{v_bound}), with $f(x)=ax$ bounding $\psi(x)$, we obtain
\begin{eqnarray}
v^{(\infty)}(x) &=& \mathbb{E}\left\{\sum_{n=1}^ke^{-r \tau^*_{n,\infty}}\psi(X_{\tau^*_{n,\infty}}^{0,x})+\sum_{n=k+1}^{\infty}e^{-r \tau^*_{n,\infty}}\psi(X_{\tau^*_{n,\infty}}^{0,x})\right\}\nonumber\\
&\leq& v^{(k)}(x)+\mathbb{E}\left\{\sum_{n=k+1}^{\infty}e^{-r \tau^*_{n,\infty}}f(X_{\tau^*_{n,\infty}}^{0,x})\right\}\nonumber\\
\label{v_inf_less_than}
&\leq & v^{(k)}(x)+aM\frac{e^{-(r-\alpha)kT}}{1-e^{-(r-\alpha)T}}.
\end{eqnarray}
In the second step we have used the fact that $(\tau^*_{n,\infty})_{n= 1}^k$ is an admissible, but not necessarily optimal, stopping rule for $v^{(k)}(x)$. With $x\leq M$ it follows from Theorem \ref{main_theorem} that $(v^{(k)})_{k\geq 1}$ is strictly increasing and bounded, and hence convergent on $[0, M]$. The uniform convergence of $v^{(k)}\to v^{(\infty)}$ on $x\in[0,M]$ then follows from  (\ref{v_inf_less_than}).\ \ \ $\square$

\section{Application to Infrastructure Investments}
\label{sec:application_to_infrastructure}
In this section, we apply our analytical results in Section \ref{sec:solution_approach} to the infrastructure investment problem and discuss a numerical implementation. Moreover, we study the sensitivity of the result with respect to the key parameters of lifetime $T$ and lead time $\nu$, as well as the process parameters.  Lastly, the proposed framework is employed to compare investment scenarios with relatively short lifetimes and lead times to a scenario with both a long lifetime and lead time. Such a comparison will reveal a critical investment cost of the short-lived scenario below which it will be competitive with the long-lived counterpart.

Recall from Section \ref{sec:problem_formulation}  the cash flow $f(X_t)$, in which we have incorporated the flexibility to temporarily suspend production to avoid a negative cash flow, namely,
\begin{equation*}
f(X_t) = \left(X_t-c\right)^+.
\end{equation*}
The   reward function associated with  this cash flow is
\begin{eqnarray}
\label{reward2}
\psi(x) = -I + \int_{\nu}^{\nu+T}\left(x\Phi(d_+(t))e^{\alpha t}-c\Phi(d_-(t))\right)e^{-rt}dt,
\end{eqnarray}
where
\begin{equation}
\label{d_s}
d_{\pm} (t)= \left[\ln\left(\frac{x}{c}\right)+\left(\alpha\pm\frac{1}{2}\sigma^2\right)t\right]/
\sigma\sqrt{t}.
\end{equation}
Since  $f (X_t )$ is bounded by $X_t$,  one realizes that there is a linear function bounding $\psi(x)$. Investigating the derivatives of $\psi(x)$ yields
\begin{equation}
\label{first_der}
\psi '(x) = \int_{\nu}^{\nu+T}\frac{d}{dx}\left(x\Phi(d_+(t))e^{\alpha t}-c\Phi(d_-(t))\right)e^{-r t}dt=\int_{\nu}^{\nu+T}\Phi(d_+(t))e^{-(r-\alpha) t}dt>0.
\end{equation}
In the second step, we have identified the derivative inside the integral as the Delta  of the European call option. From (\ref{reward2})-(\ref{first_der}) we see that $\psi(x)$ is continuous and increasing in $x$. Furthermore, since $\lim_{x\to 0}\psi(x)=-I<0$ there exists a unique break-even point $x_0$. Thus, $\psi(x)$ satisfies all the conditions in the definition of a reward function presented in Section \ref{sec:opt_stopping_rule}. By differentiation, we find that
\begin{equation*}
\frac{d^2}{dx^2}[\Lambda\psi(x)] =
\int_{\nu}^{\nu+T}\frac{\phi(d_+(t))}{x\sigma^2t}\left((\gamma-1)\sigma\sqrt{t}+d_+(t)\right)e^{-(r-\alpha) t}dt,
\end{equation*}
where $\phi(x)$ is the density function of the normal distribution.
It can be seen that $\Lambda\psi(x)$ is convex for all $x\geq x'$, where $x'=c\exp\left[-\left(\alpha+\gamma\sigma^2-\frac{1}{2}\sigma^2\right)\nu\right]$. In turn, if $x_0\geq x'$, which is ensured by imposing a lower bound on $I$, then $\Lambda\psi(x)$ is convex on $(x_0,\infty)$, and consequently, all the conditions for Theorem \ref{main_theorem} are satisfied and the algorithm outlined in Corollary \ref{algorithm_prop} can be applied.

If the operational flexibility to suspend operation is removed then the expected cash flow can be readily integrated and the reward function $\psi_0(x)$  in this case would be
\[
\psi_0(x) = -I+\int_{\nu}^{\nu+T}e^{-r t}\,\mathbb{E}\left\{\left(X_t^{0,x}-c\right)\right\}\,dt=a(\nu,T)x+b(\nu,T),\]
where $a(\nu,T)$ and $b(\nu,T)$ are functions of the lead-time and lifetime only. That is, without the operational flexibility the reward function $\psi_0(x)$ is clearly linear in $x$, rendering $\Lambda \psi_0(x)$ convex on $\mathbb{R}^+$, and thus Theorem 1 is also applicable to this case.

\subsection{Numerical Implementation}
\label{sec:implementation}
The expressions for $u^{(k)}(x)$ and $v^{(k)}(x)$ in \eqref{u_def} and \eqref{value_defined_through_u} above form the basis of our numerical algorithm, whereby $u^{(k)}(x)$,  $v^{(k)}(x)$, and $x_k^*$ are computed iteratively for $k=1, 2, 3, \ldots$
The calculations  below were carried out on a grid of 500 points regularly spaced between 0 and $x_{\max}$, where the latter was determined by the process and model paramenters. The computation of the expectation in (\ref{u_calc}) is complicated by the fact that $u^{(k-1)}(x)$ is not bounded on $\mathbb{R}^+$. However, since our choice of reward function, $\psi(x)$ in (\ref{reward2}), is approximately affine for large $x$, so are $\Lambda\psi(x)$ and $u^{(k-1)}(x)$. Rather than truncating the distribution we can instead linearly extrapolate $u^{(k-1)}(x)$ outside the given grid. Since $(x_k^*)_{k\geq 1}$ is decreasing, the necessary range of the grid depends mainly on $x^*_1$. Specifically, $x_{\max}$ was chosen to be the upper bound on a two-sided 99.9\% confidence interval around $x_1^*$,
\begin{equation*}
x_{\max} = \exp\left(\ln(x_1^*)+\left(\alpha-\frac{1}{2}\sigma^2\right)T+3.29\sigma\sqrt{T}\right).
\end{equation*}
This choice of $x_{\max}$, together with the number of gridpoints, was an acceptable compromise between having a large enough grid to ensure a linear behavior of $u^{(k-1)}(x)$ without undue computational complexity. When comparing different scenarios, i.e. different values for $T,\nu$ and $I$, the largest grid was used throughout.

Using a trapezoidal method, the expectation in (\ref{u_calc}) was calculated as
\begin{equation}
\label{exp_calc}
\mathbb{E}\left\{u^{(k-1)}(X_T^{0,x})\right\} = \int_0^{x_{\max}}u^{(k-1)}(z)g(z;x,T,\alpha,\sigma)dz+\int_{x_{\max}}^{\hat{x}}(kz+m)g(z;x,T,\alpha,\sigma)dz,
\end{equation}
where $kz+m$ is the linear extrapolation of $u^{(k-1)}$ for $x>x_{\max}$ and where $g(z;x,T,\alpha,\sigma)$ is the density function of the lognormal distribution. The upper limit $\hat{x}$ in (\ref{exp_calc}) was chosen so that the support of the distribution contained a two-sided confidence interval around $x$, for every $x\leq x_{\max}$.

The boundary points $x_k^*$ were found using a simple bisection method on the convex function in (\ref{boundary_point}). The calculations of $u^{(k)}(x)$, and therefore also of $x^*_k$ and $v^{(k)}(x)$, were terminated once a tolerance $\varepsilon$, defined by
\begin{equation}
\varepsilon = \frac{|u^{(k)}-u^{(k-1)}|}{|u^{(k)}|}, \label{tolerance}
\end{equation}
had reached $\varepsilon\leq 10^{-3}$. Such a tolerance yields a solution $v^{(k)}(x)$ that is within $0.1$\% of $v^{(\infty)}(x)$, an accuracy that is likely good enough considering reasonable errors in the estimation of the underlying parameters. In the results below we denote by $x_{\infty}^*$ and $v^{(\infty)}(x)$ the stopping boundary and the value function respectively at the termination of the algorithm according to the tolerance in \eqref{tolerance}.

\begin{figure}[h!]
\includegraphics[width=3.1in]{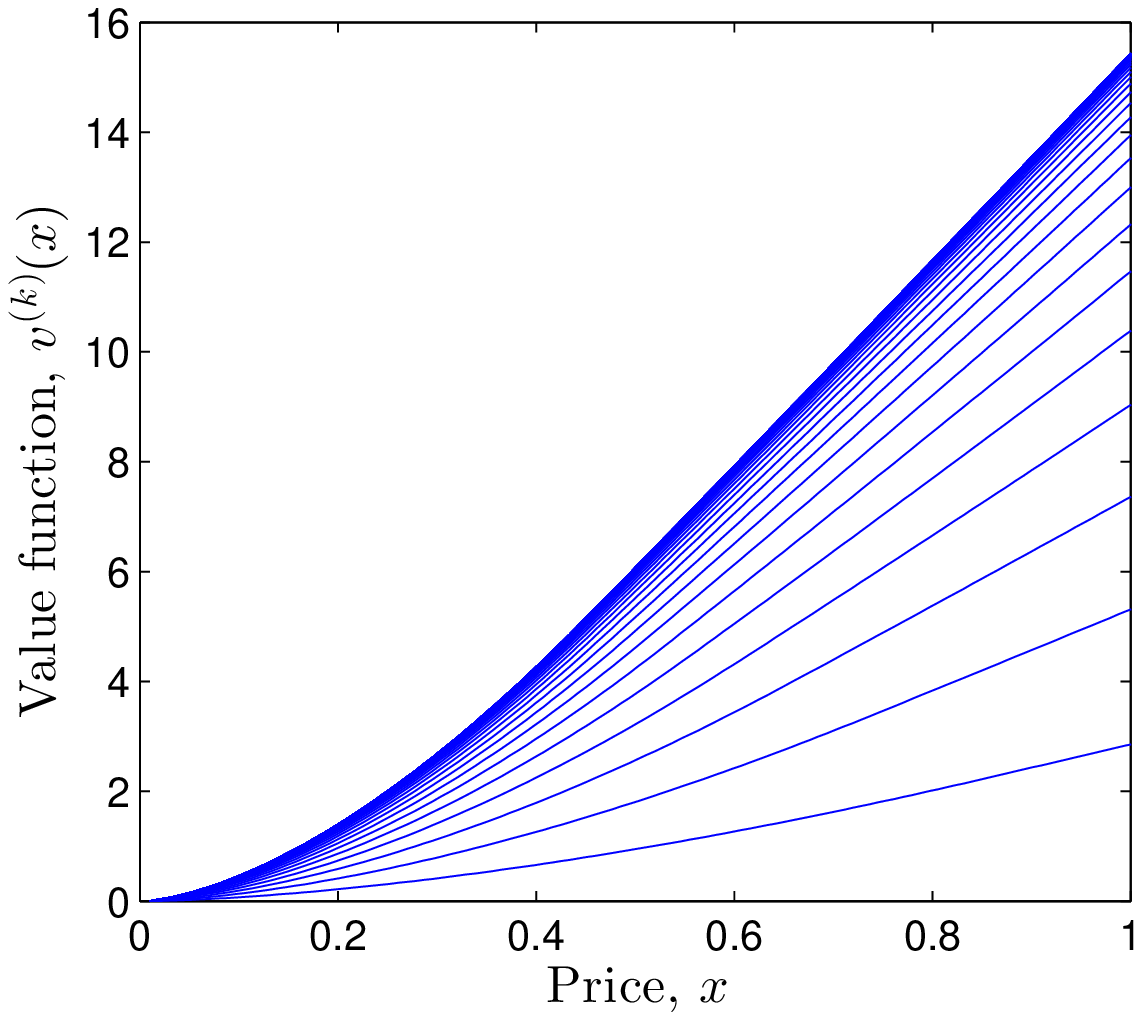}
\includegraphics[width=3.1in]{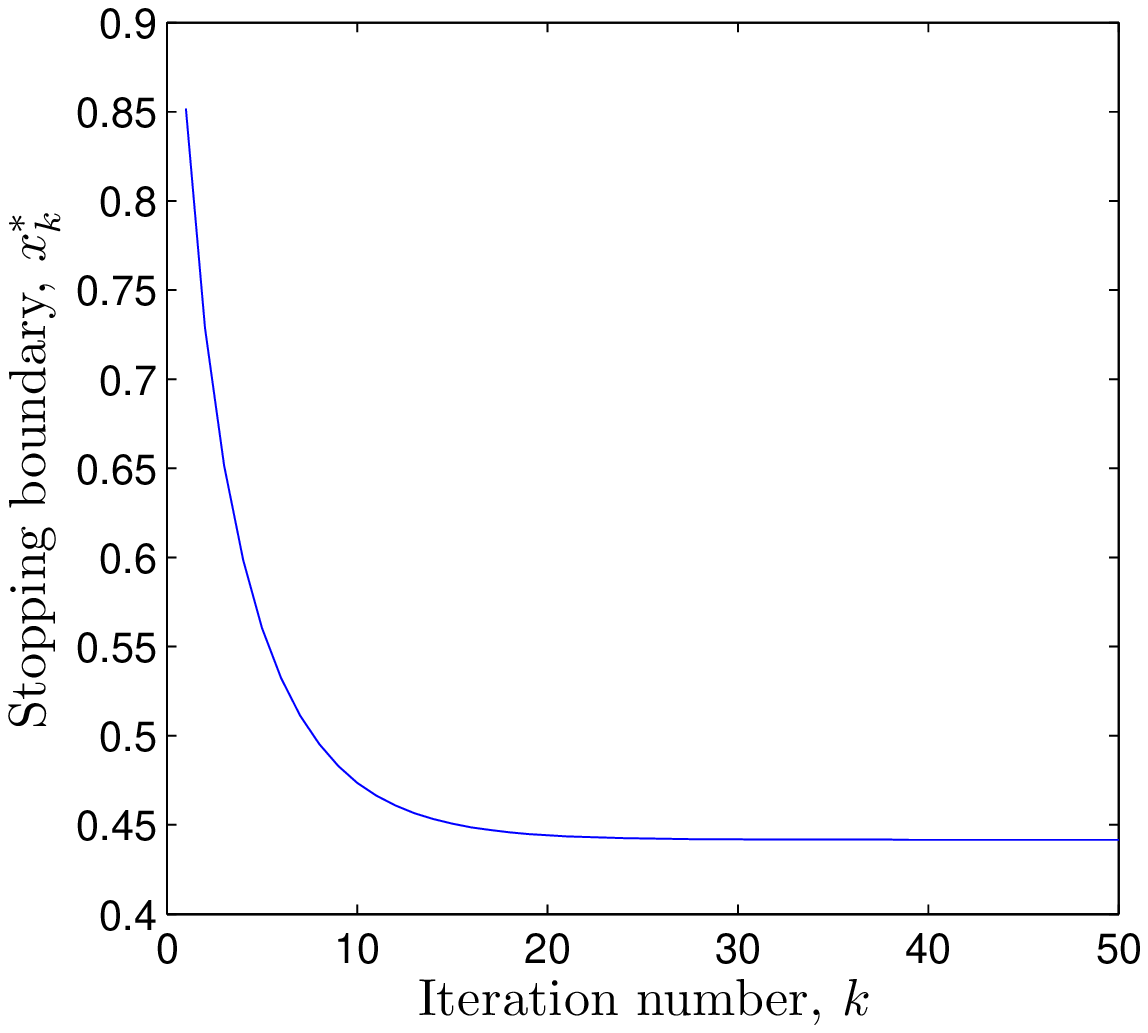}
\caption{\small{(Left) Convergence of the value function, $v^{(k)}(x)$,  for $k = 1,\dots,50$. (Right)  The stopping boundary $x_k^*$ decreases rapidly over iterations.}}
\label{fig:convergence}
\end{figure}

Figure \ref{fig:convergence} demonstrates the convergence of the algorithm for the value function $v^{(k)}(x)$ and the stopping boundary $x^*_k$ respectively. Default parameter values used in the calculations below are given in Table \ref{table:parameters}. The choice of fixed cost parameters ($I$ and $c$), or their ratio $I/c=10$, stems from observations in pertinent industries. Recent studies on energy investments  and  commodities report  price volatilities in the neighborhood of 20\% though they can occasionally experience short-term  drastic increases (see \cite{Westner201231,Geman2009576}). Therefore,  setting $\sigma=20\%$  corresponds to a relatively conservative choice as lower volatility favors the status quo. However, we will also examine a relatively high volatility ($\sigma = 40\%$) scenario in Figure \ref{fig:operational_flex}.   Additionally, the use of an effective discount rate $(r-\alpha)$ of 5\% is also in line with examples found in the literature in related fields, such as \cite{Frayer200140}. Overall, the choices of parameter values are, albeit reasonable, mainly for illustrative purposes.

According to  Theorem \ref{main_theorem}, $\left(v^{(k)}(x)\right)_{k\geq 1}$ is strictly increasing and the sequence of stopping boundaries $\left(x^*_k\right)_{k\geq 1}$ is strictly decreasing. In Figure \ref{fig:convergence} (left), we see that the value function increases monotonically with  each iteration.
 After 50 iterations the tolerance $\varepsilon$, defined above, was less than $10^{-5}$. Moreover, we notice that the value function appears linear for large $x$.  In Figure \ref{fig:convergence} (right), the stopping boundary $x_k^*$ decreases rapidly from 0.85 to 0.44 and we mention that the break-even point was $x_0=0.33$. The convergence is clear even after 20 iterations. In particular, the  stopping boundary $x^*_1 =0.85$ (see \eqref{Lambda})  from the first iteration helps us define the upper bound $x_{\max}$ of the grid.

The stopping boundary  $x^*_k$  is the price level at or above which the first investment should be made, given the option to  make $k-1$ more  investments later, each separated in time by at least the lifetime $T$. Since  firms rarely have an imposed number of investment renewals, the boundary  $x^*_{\infty}$ is of primary interest. Contrasting the value of $x^*_1$ to $x^*_{\infty}$ reveals the impact of including future investment options on current investment decision. In Figure \ref{fig:convergence},   the number of exercises (iterations) is $k=50$, which together with the lifetime of $T=5$ years,  gives a time horizon of more than $250$ years. Such a time horizon is practically infinite under reasonable circumstances. On the other hand,  with a large $k$, the number of  remaining investment opportunities should have a smaller impact on the first investment timing. This is evidenced by  the convergence of  $x^*_k$ to a constant level as $k$ increases.

 From our numerical tests, we find that the refraction time (lifetime) $T$ strongly  influences the speed of convergence. This is intuitive due to   the discount factor $e^{-rT}$ in the  definition of $u^{(k)}(x)$ in (\ref{u_calc}), reducing the differences over iterations.    On the other hand, the lead time $\nu$ has much less bearing on the rate of convergence since it only affects the first investment timing.  We will further examine the impacts of lifetime and lead time in the next subsection.

\begin{table}[t!]
\centering
\begin{tabular}{l|l|l}
\hline
Description & Parameter & Value\\
\hline
Lifetime & $T$ & 5\\
Lead time & $\nu$ & 1\\
Investment cost & $I$ & 1\\
Operational cost & $c$ & 0.1\\
Discount rate & $r$ & 10\%\\
Drift rate & $\alpha$ & 5\%\\
Volatility & $\sigma$ & 20\%\\
\hline
\end{tabular}
\caption{\small{Default parameter values used in the calculations. }}
\label{table:parameters}
\end{table}

\subsection{Sensitivity Analysis}\label{sec:sensitivity}
 We will first investigate the sensitivity of $x_{\infty}^*$ with respect to the process parameters $\alpha$ and $\sigma$. Later we investigate the sensitivity of both $x_{\infty}^*$ and the value function $v^{(\infty)}(x)$ with respect to the main model parameters of lifetime $T$ and lead time $\nu$.

\begin{figure}[th]
\includegraphics[width=3.1in]{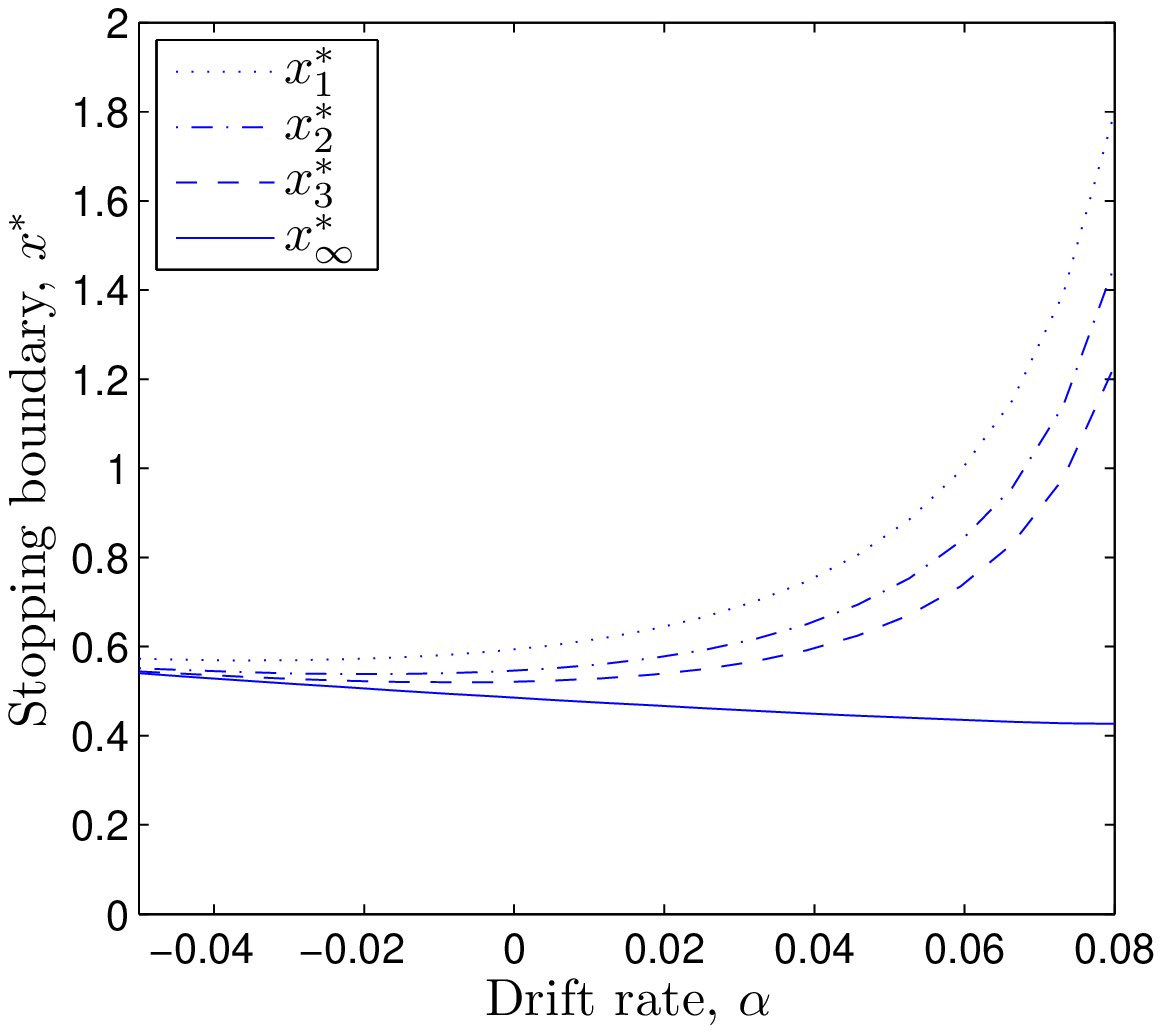}
\includegraphics[width=3.1in]{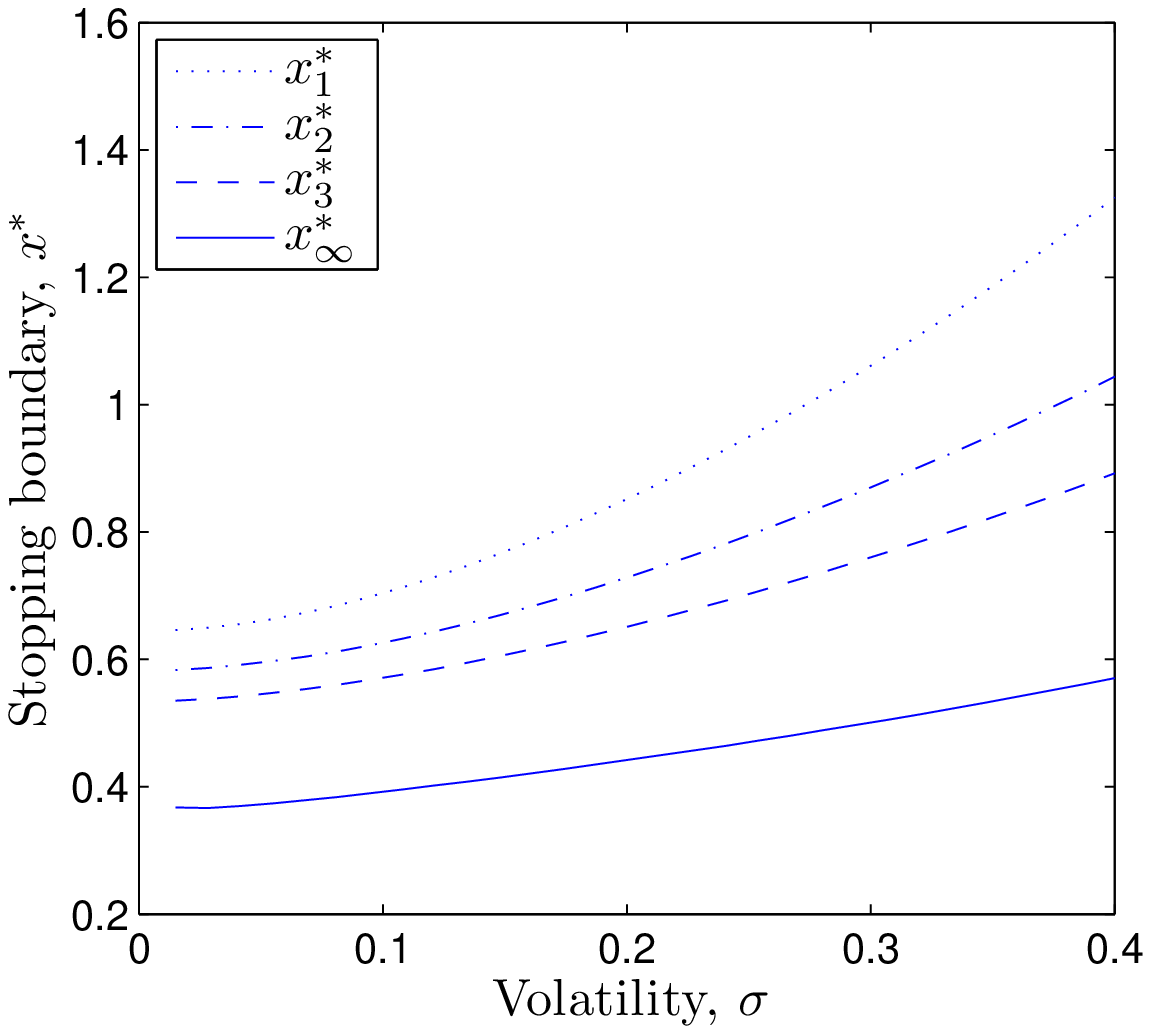}
\caption{\small{Sensitivity of the optimal stopping boundaries $x_k^*$ w.r.t. the drift rate $\alpha$ (Left), and w.r.t. the volatility $\sigma$ (Right).}}
\label{fig:parameter_sensitivity}
\end{figure}

Low values of the drift rate $\alpha$ corresponds to a higher `effective' discount rate $r-\alpha$, sometimes called the convenience yield. This decreases the present value of future investment, which explains the minor difference between $x_1^*$ and $x_{\infty}^*$ for small $\alpha$, see Figure \ref{fig:parameter_sensitivity} (left). Conversely, a higher drift rate of the underlying enhances the value of future investments. This drives a greater wedge between the stopping boundary $x_1^*$ of a single investment compared to the stopping boundary $x_{\infty}^*$ of multiple consecutive investments for large $\alpha$. Consequently, this emphasizes the importance of including future investments in current decisions in environments with a high drift rate. Note that the same result is to be expected if decreasing the discount rate $r$, still with the condition that $r-\alpha>0$. In the figure,  we observe that as $\alpha$ approaches $r$ (10\%) the exercise boundaries $x_k^*$, $k=1,2,3$, increases rapidly. This is intuitive because theoretically the optimal exercise boundary would be infinite for $\alpha \ge r$. The same phenomenon also occurs for $x^*_\infty$ when $\alpha $ is very close to $r$.

In Figure \ref{fig:parameter_sensitivity} (right), we see that the optimal exercise boundary $x^*_\infty$ increases with volatility $\sigma$.  This suggests that in a more volatile environment the firm will demand a higher output price level in order to enter the market. %, and hence delaying the investment decision.  
We observe that the increasing pattern holds for finite $k$, $k=1,2,3$, as well as for the infinite case.

\begin{figure}[ht]
 \includegraphics[width=3in]{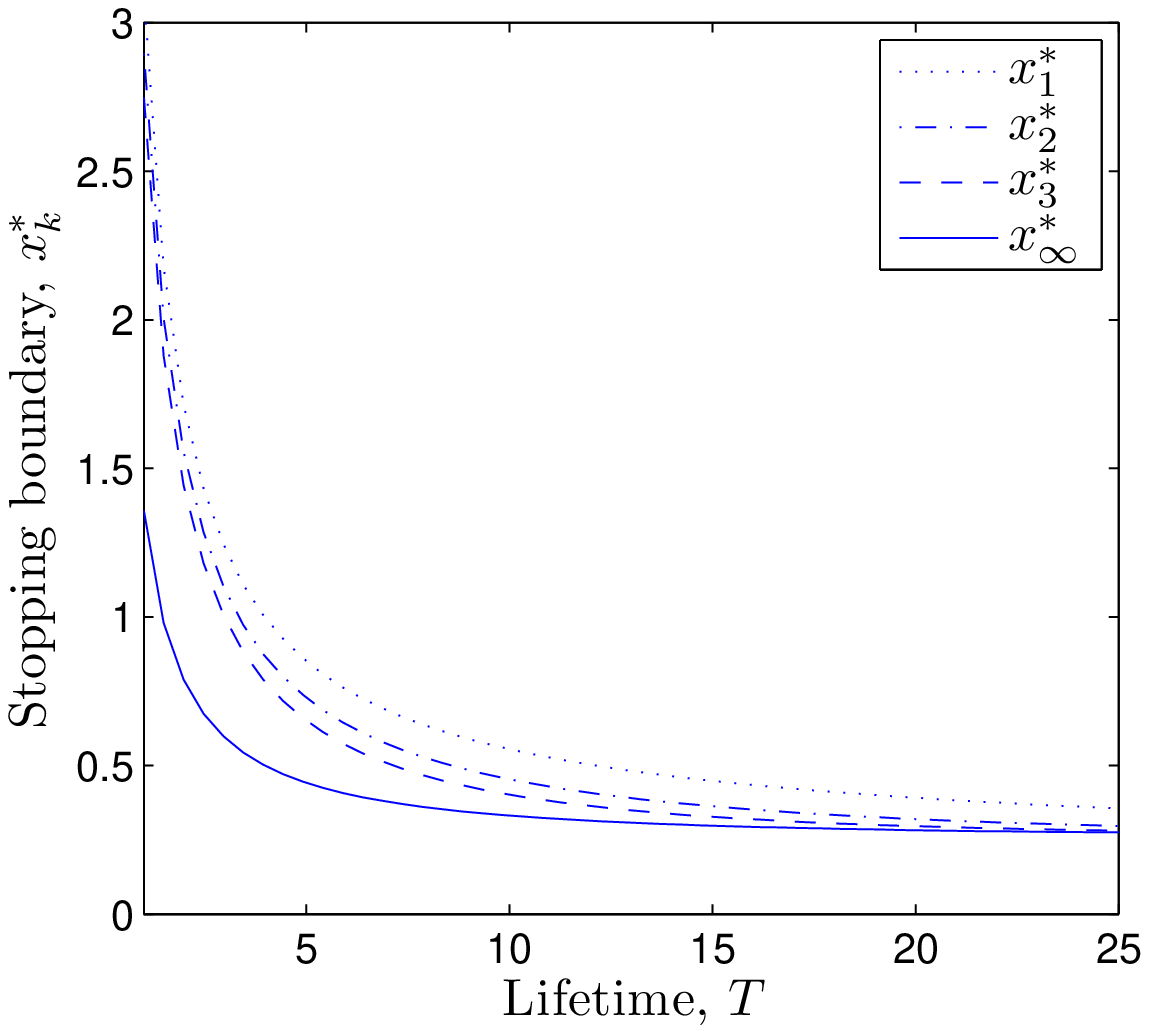}\includegraphics[width=3in]{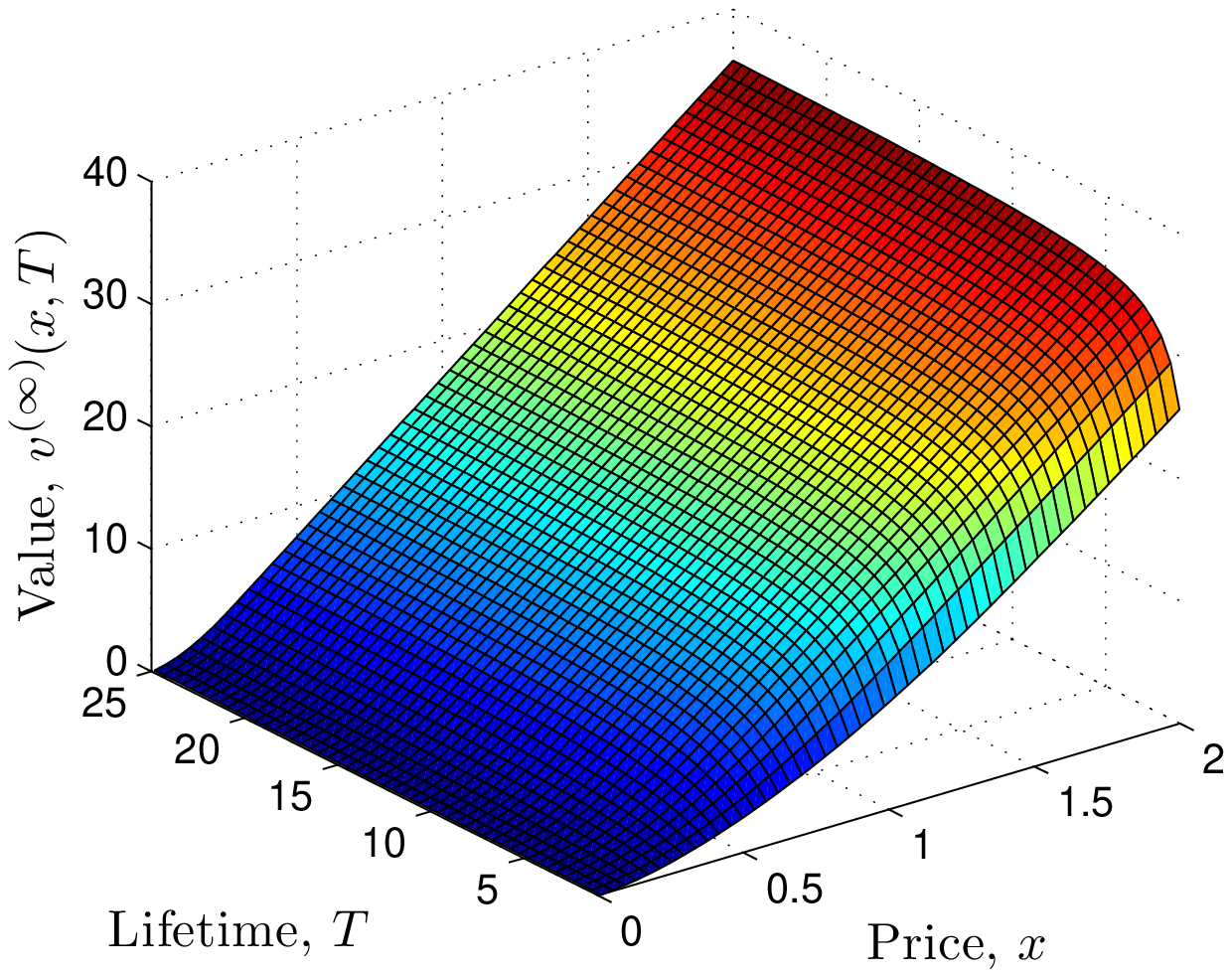}
\caption{\small{(Left) Optimal stopping boundaries $x_k^*$ decrease as lifetime $T$ increases.  (Right) The value $v^{(\infty)}(x,T)$ is increasing with respect to the underlying price levels $x$ and lifetimes $T$.}}
\label{fig:lifetime_value}
\end{figure}

Now, we turn to examine the impact of lifetime $T$, the parameter of main interest, on the stopping boundaries $x_k^*$ and value $v^{(k)}(x)$,  with  the same fixed cost $I=1$.  First, in Figure \ref{fig:lifetime_value}, we observe  a minor difference between $x^*_1$ and $x^*_{\infty}$ for long lifetimes (for $T \ge 20$). Intuitively, the incremental  value of an additional investment 20 years or more from now  is minimal  due to the discount factor $e^{-r T}$ in (\ref{u_calc}) and (\ref{boundary_point}). Therefore, for very long lifetimes,  future investment decisions  will   not significantly influence the current decision to invest. However, for very short lifetimes, $T<2$, there is a substantial difference between the optimal exercise levels with one and infinite investment opportunities (i.e.   $x^*_1$ vs. $x^*_{\infty}$ for small $T$).   In an intermediate regime, with lifetimes of 5 to 15 years, we observe that including only one or two future investment decisions will significantly affect the decision regarding the first investment.  In both finite and infinite cases, the optimal exercise boundary decays rapidly with respect to lifetime. For instance, $x^*_{\infty}$  is almost flat for lifetime of 10 years or longer.

Considering the value $v^{(\infty)}(x)$ of the multiple investment scenario for different lifetimes we observe a trend similar to the one regarding stopping boundary. The marginal impact of adding one year of life to short-lived capital is significant. However, this impact drastically decreases for longer-lived capital. For instance, as seen in Figure \ref{fig:lifetime_value}, the value of multiple investments in capital with a 25 year lifespan is virtually the same as capital with a life of 5 years, at the same investment cost $I$ and the same lead time $\nu$. Even though the investment cost is the same, the increased flexibility in timing future investments in the scenario with the shorter lifetime makes them almost equally attractive.

\begin{figure}[ht]
\centering
\includegraphics[width=3in]{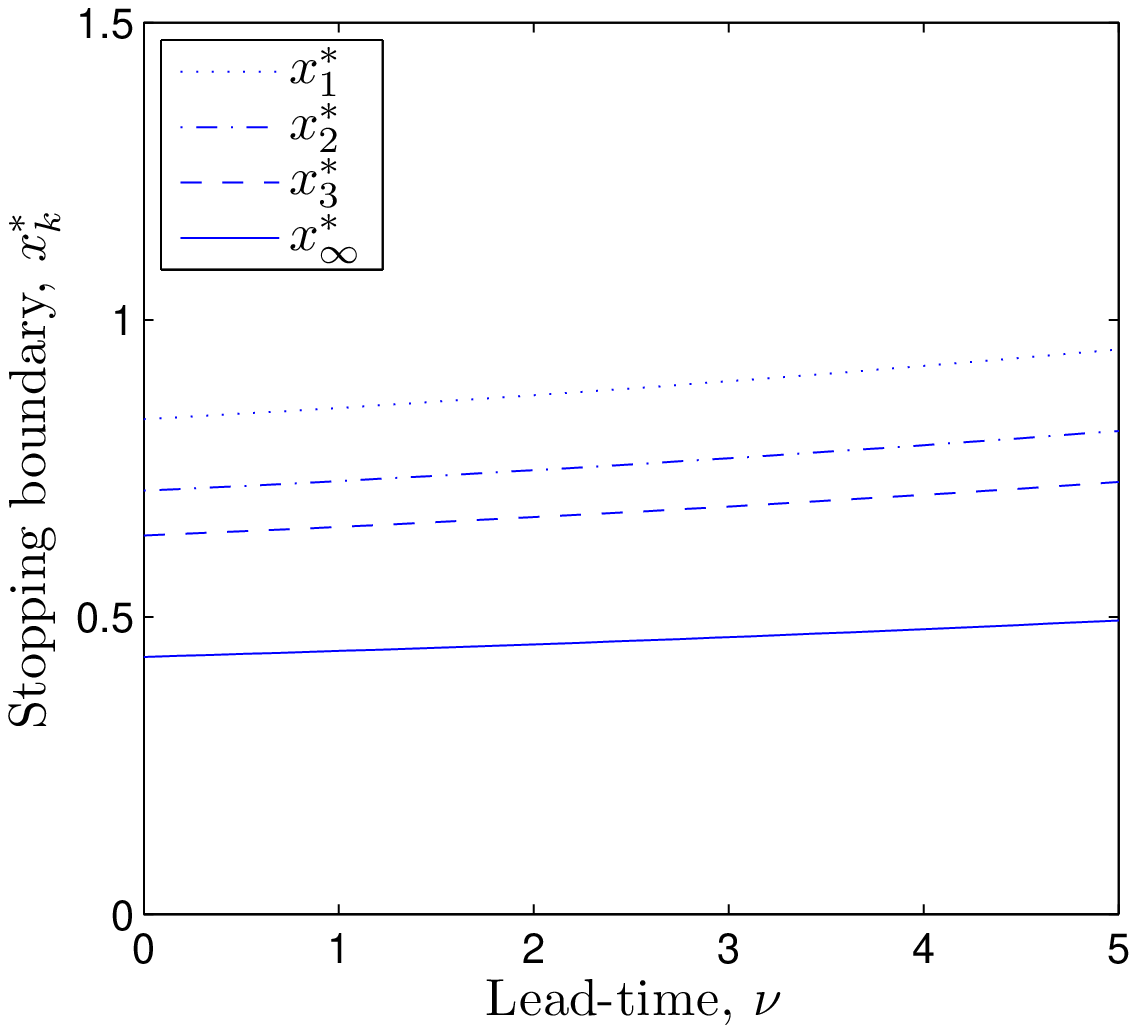}
\includegraphics[width=3in]{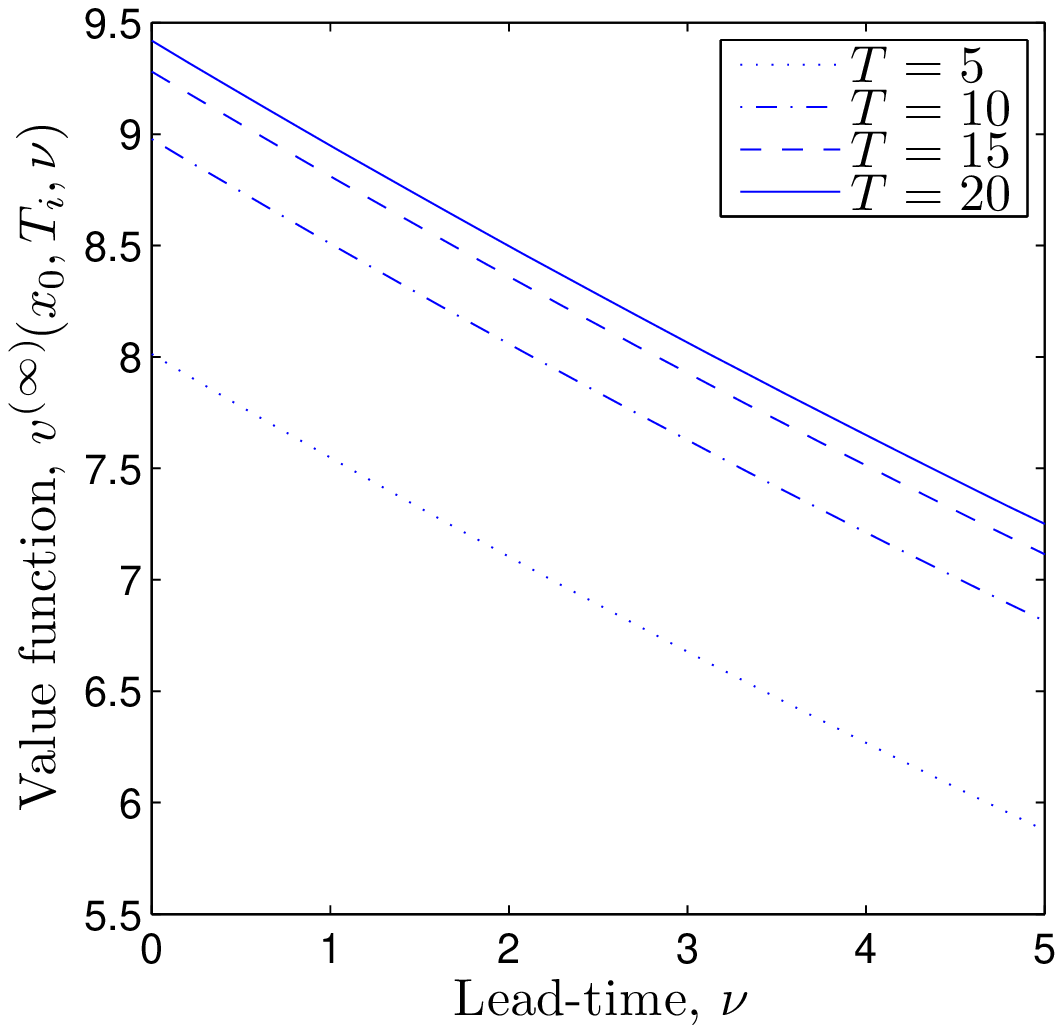}
\caption{\small{(Left) Optimal stopping boundaries $x^*_k$ for different lead times $\nu$ with the same fixed cost $I=1$. (Right)
The value $v^{(\infty)}(x_0,T_i,\nu)$ evaluated for different lead times $\nu$ at $x_0 = 0.5$, and for $T_i = 5, 10, 15,20$. }}
\label{fig:leadtime}
\end{figure}

 As seen in Figure \ref{fig:leadtime} (left), the lead time does not substantially influence $x_{\infty}^*$, nor the difference between $x_1^*$ and $x_{\infty}^*$ at fixed $\nu$. In other words, varying lead times (within the given range) will barely affect the firm's decision to invest.  This is expected since seamless consecutive investments are possible regardless of the length of lead time. On the other hand, a longer lead time  delays revenue generation  relative to  the outlay of the investment cost $I$. Increasing lead time therefore decreases the net present value of every future investment. This explains the decreasing trend of  the value function  $v^{(k)}$  with respect to lead time in Figure \ref{fig:leadtime} (right).  In summary,  the firm will realize a lower value with longer lead times of the investments, even though   the corresponding  exercise boundary changes only marginally. Displaying the value $v^{(\infty)}$ at different lead times $\nu$ and at a fixed price level but for several values of the lifetime $T$ again shows the diminishing returns of adding lifetime to capital.

From Section \ref{sec:problem_formulation}, we know that the reward function $\psi$ can be interpreted as the sum (integral) of European call options on the uncertain output price $X$, with strike $c$ and  maturities ranging over the lifetime. Since increasing the strike price of a European call decreases its value, the cost parameter $c$ has the same effects on  the value $v^{(k)}$, which in turn  raises the stopping boundary $x^*_k$. By similar reasoning, the same holds for $I$.

\begin{figure}[ht]
\centering
\includegraphics[width=3in]{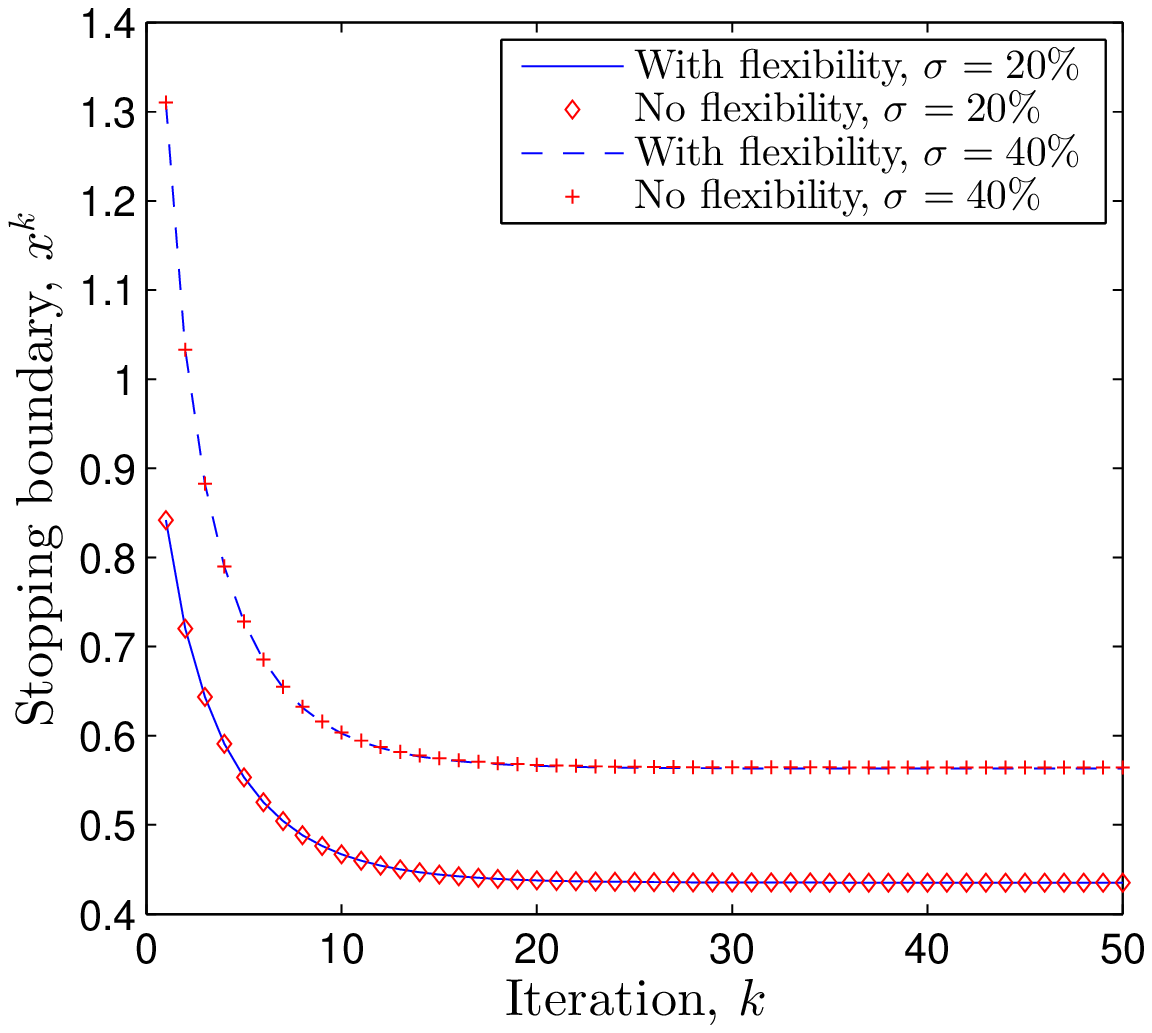}
\includegraphics[width=3in]{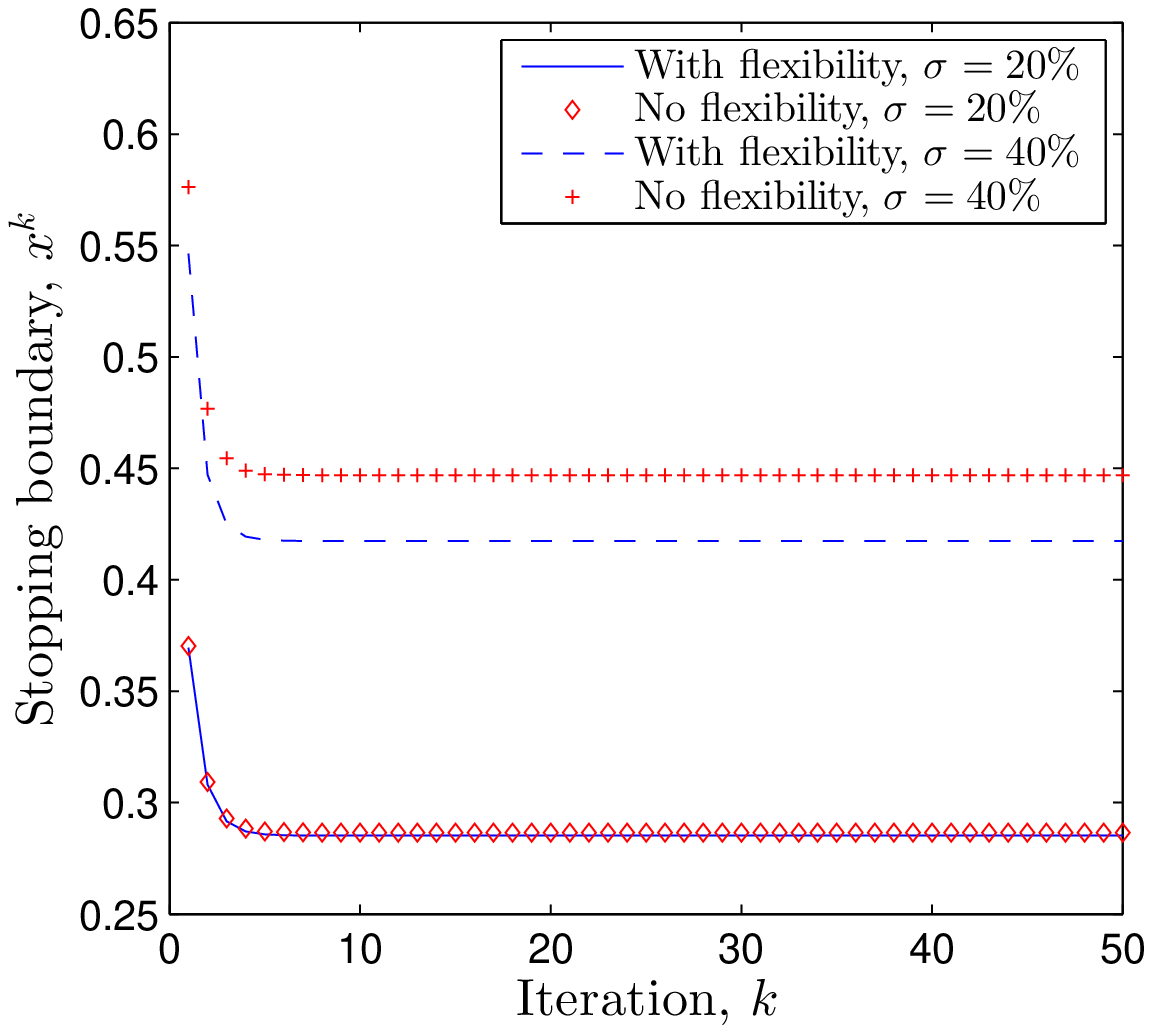}
\caption{\small{(Left) Stopping boundary for a short-lived scenario where capital has a lifetime $T=5$ years and a lead-time $\nu=0.5$ years. (Right) Stopping boundary for a long-lived scenario where capital has a lifetime  $T=25$ years and lead-time $\nu=5$ years. }}
\label{fig:operational_flex}
\end{figure}

Lastly, we examine the value of the option to  temporarily suspend production for investments with  short and long lifetimes. Two scenarios are compared in Figure \ref{fig:operational_flex}, short-lived investments ($T=5$, $\nu=0.5$) and long-lived investments ($T=25$, $\nu=5$) at two levels of volatility $\sigma=20\%$ and $40\%$ respectively.  Note that in both cases an investment cost of $I=1$ was used, which explains a relatively higher stopping boundary for the short-lived capital.  In this example, the operational flexibility  appears to barely  affect the investment decision in the short-lived scenario regardless of volatility. For low volatility the same indifference appears in the long-lived scenario as well,  Figure \ref{fig:operational_flex} (right). However, in contrast to the short-lived scenario, the operational option leads to  earlier investment timing  under  high  volatility  in the long-lived scenario.  Comparing the left and right panels of Figure \ref{fig:operational_flex}, we see that a   shorter lifespan yields higher thresholds for future investments.

 This operational flexibility feature is typically associated with competitive and unregulated markets, whereas many of the relevant industries for this paper traditionally have operated in regulated environments. Moreover, it could be argued that the conventional `bigger-is-better' paradigm has historically invited regulations on the market since monopolies are often observed in  industries with a large production scale.  However, it is conceivable that these conditions may change as new technological advances allows for more competitive  distributed small-scale production models.  

 As an example, these changes are occurring and subsequently making operational flexibility available  in the electric power generation. In the United States,  more than half of the states have moved towards deregulated power markets. While power producers face many different costs the dispatch order of the various available generators in a given market is to zeroth order a function of operational cost. That is, it is up to the individual producers to decide a price point where operation commences.   It should be noted that the interpretation of a unit not operating is dependent on the time scales involved. For instance, hour-to-hour decisions typically implies putting generators into an synchronized  reserve capacity called  \textit{spinning reserves},  whereas seasonal interruptions correspond to complete \textit{cold stop}. Some technologies, like nuclear power, currently have a relatively  low operational cost and are thus ahead in the dispatch order to meet demand (also called base load  generation).  In general, the value of  the  operational flexibility  option is expected to be  higher     if the production involves   lowers costs of suspension and resumption.  Lastly, we remark that our standing  assumption  that  the firm acts as a price taker would have to be revisited if  the resulting change  in supply should  significantly  impact   the market price.

\subsection{Critical Investment Cost}
\label{sec:big_v_small}
Many, if not all, of the fundamental process industries, such as e.g. energy, water and petrochemicals, have followed the trend of `bigger-is-better' over the past century. Experience from a given industry gives us the investment cost (per unit of capacity) $I_{large}$, as well as the lifetime $T_{large}$ and lead time $\nu_{large}$ of large-scale capital in the current paradigm. Transitioning to a paradigm of small-scale, mass-produced, and modular equipment will give rise to a new parameter ensemble, $\{I_{small}, T_{small},\nu_{small}\}$, where we  a priori  only infer shorter lifetimes and lead times. By considering an infinite time horizon, we can use the framework in this paper to find the critical investment cost $I_{crit}$ that would render a small-scale approach competitive. That is, for a given reference ensemble $\{I_{large}, T_{large},\nu_{large}\}$,  we find for every choice of $T_{small}$ and $\nu_{small}$   an $I_{crit}$ such that $v^{(\infty)}_{small}\geq v^{(\infty)}_{large}$, provided that $I_{small}\leq I_{crit}$.

The observed historical trend of increasing unit sizes seemingly suggests that \emph{total} cost decrease with size. Nevertheless, we will here let the \emph{operational} cost $c$ be independent of the parameters $I, T$ and $\nu$. To motivate this setting for our examples below, we consider  the total capacity to be installed at a single location, so  ancillary costs, such as transportation costs of inputs and outputs to and from the plant, administrative costs, cost of security etc., ought not depend on the granularity of the hardware inside this black box. This leaves primarily two potentially size-dependent factors that can affect operational costs; labor and efficiency.
Historically, the necessary amount of operational labor increases with the number of individual units employed. Operating fewer and larger units therefore increases labor productivity and lowers labor cost per unit output. However, advances in automation make it possible today to decouple the amount of required labor from the number of individual units at reasonable cost. As to efficiency,  from a physical perspective one can in some cases claim that larger units have lower dissipative losses (e.g. through friction and unwanted heat transfer) and hence are more efficient and therefore less costly to operate. On the other hand, it is not always clear how important these effects are. For instance, a statistical analysis on the size-dependency of cost in four different electricity generating technologies showed that once labor cost is removed, unit size is not a significant factor affecting total operational cost \citep{Dahlgren}.

As an illustration, we can compare a single-cycle thermal power plant to an internal combustion engine, both performing fundamentally the same task of converting chemical energy into mechanical work. Under reasonable circumstances, they can do so at comparable efficiencies. The car engine is mass-produced on the order of days. The power plant, on the other hand, is typically not ready for operation until several years have passed since the decision to invest was made. Moreover, the power plant is designed to last for decades while the car engine presumably will have a lifetime on the order of years under constant operation. How much would one be willing to pay for an engine that is fully automated, retrofitted to run on the same fuel and equipped with a generator to produce electricity? Such a mini-power plant can reasonably be assumed to incur similar levels of operational costs per kWh produced as its large-scale counterpart. This leaves investment cost, lifetime and lead time as the main distinguishing features from the large-scale power plant.

We start by analyzing the reward function in (\ref{reward2}) for large $x$. It can be seen that $\Phi(d_{\pm})\approx 1$ for $x\gg c$, and therefore, $\psi(x)$ is asymptotically affine with
\begin{eqnarray}
\psi(x) &\approx& -I+\int_{\nu}^{\nu+T}\left(xe^{\alpha t}-c\right)e^{-r t}dt\nonumber\\
\label{net_value_2}
&=&\frac{e^{-(r-\alpha)\nu}}{r-\alpha}(1-e^{-(r-\alpha) T})x-\left(I+\frac{e^{-r\nu}}{r}(1-e^{-r T})c\right)=ax-b.
\end{eqnarray}
For large enough values of $x$ the function $u^{(k)}$ is affine as well and
\begin{eqnarray}
u^{(k)}(x)&=&\Lambda\psi(x) +e^{-r T}\mathbb{E}\left\{u^{(k-1)}(X_T^{0,x})\right\}\approx\Lambda\psi(x) +e^{-r T}u^{(k-1)}\left(\mathbb{E}\left\{X_T^{0,x}\right\}\right)\nonumber\\
&=&\sum_{i=0}^k\left(e^{-(r-\alpha)iT}(\gamma-1)ax-e^{-r iT}\gamma b\right),\nonumber
\end{eqnarray}
where $\mathbb{E}\left\{X_T^{0,x}\right\} = xe^{\alpha T}$. Using (\ref{net_value_2}) to substitute for $a$, we obtain for large $x$:
\begin{eqnarray}
u^{(\infty)}(x)&\approx&\frac{e^{-(r-\alpha)\nu}}{r-\alpha}x-\frac{\gamma b}{1-e^{-r T}}\,,\nonumber
\end{eqnarray}
where the slope is independent of the lifetime $T$ and decreasing in the lead time $\nu$. From the definition of $u^{(k)}$ in (\ref{u_def}) it follows that the same properties  can be ascribed to the value function $v^{(k)}(x;T,\nu)$ for large $x\gg c$. Consequently, comparing different scenarios, i.e. different $T$, $\nu$ and $I$, the scenario with the shortest lead time will always have the higher value for large enough values of $x$.

Now we  compare the value of small-scale capital, here characterized  by $T_{small}\leq 5$ years, $\nu_{small}\leq 3$ years, against a benchmark of $T_{large}=25$ years, $\nu_{large}=5$ years, representing traditional large-scale capital. Furthermore, as a point of reference we set $I_{large}=1$, which together with the constant operational cost $c=0.1$ (assumed the same for every choice of $I$, $T$ and $\nu$) provides a relative monetary scale.

\begin{figure}[th]
\begin{center}
\includegraphics[width=3in]{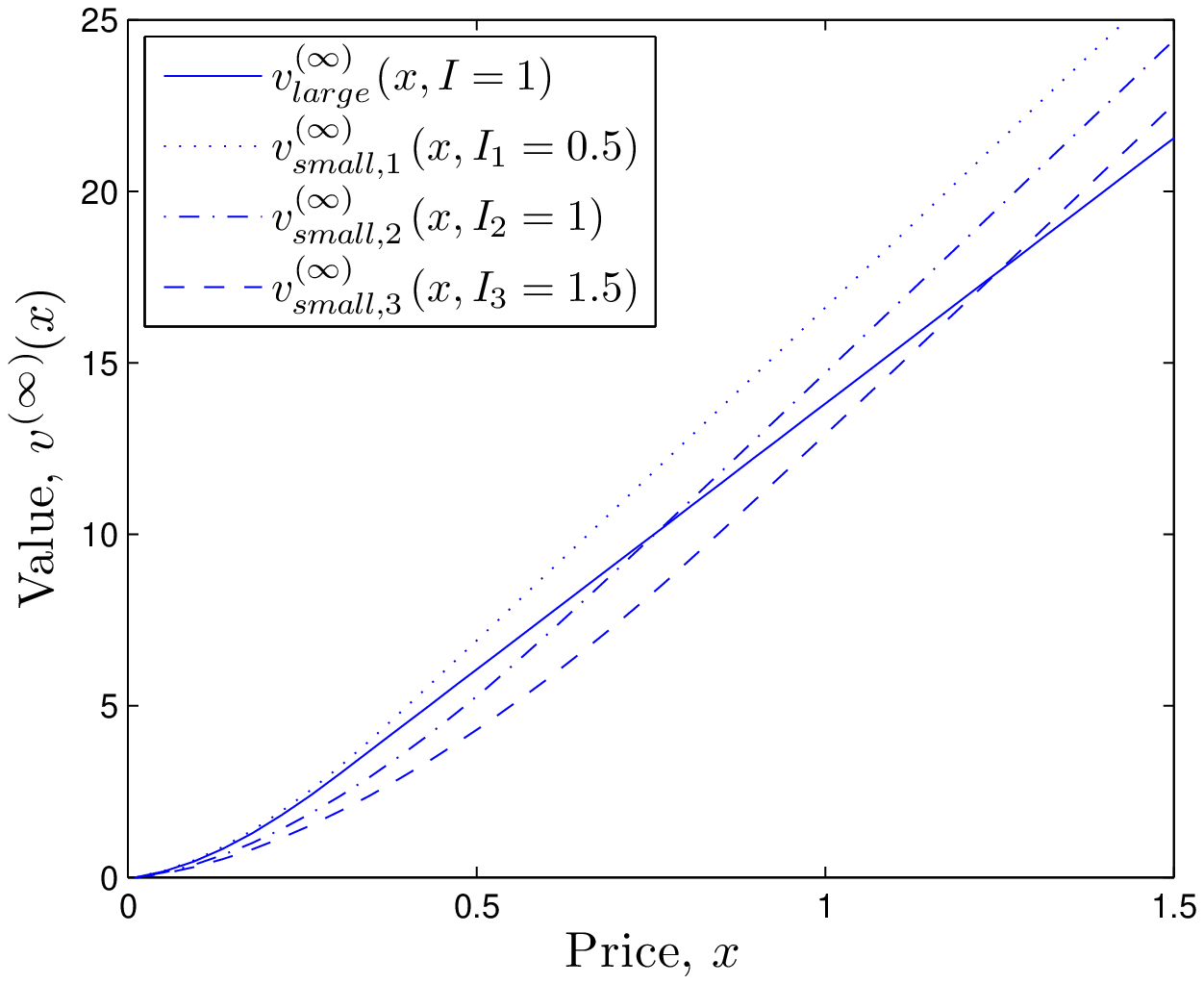}
\includegraphics[width=3in]{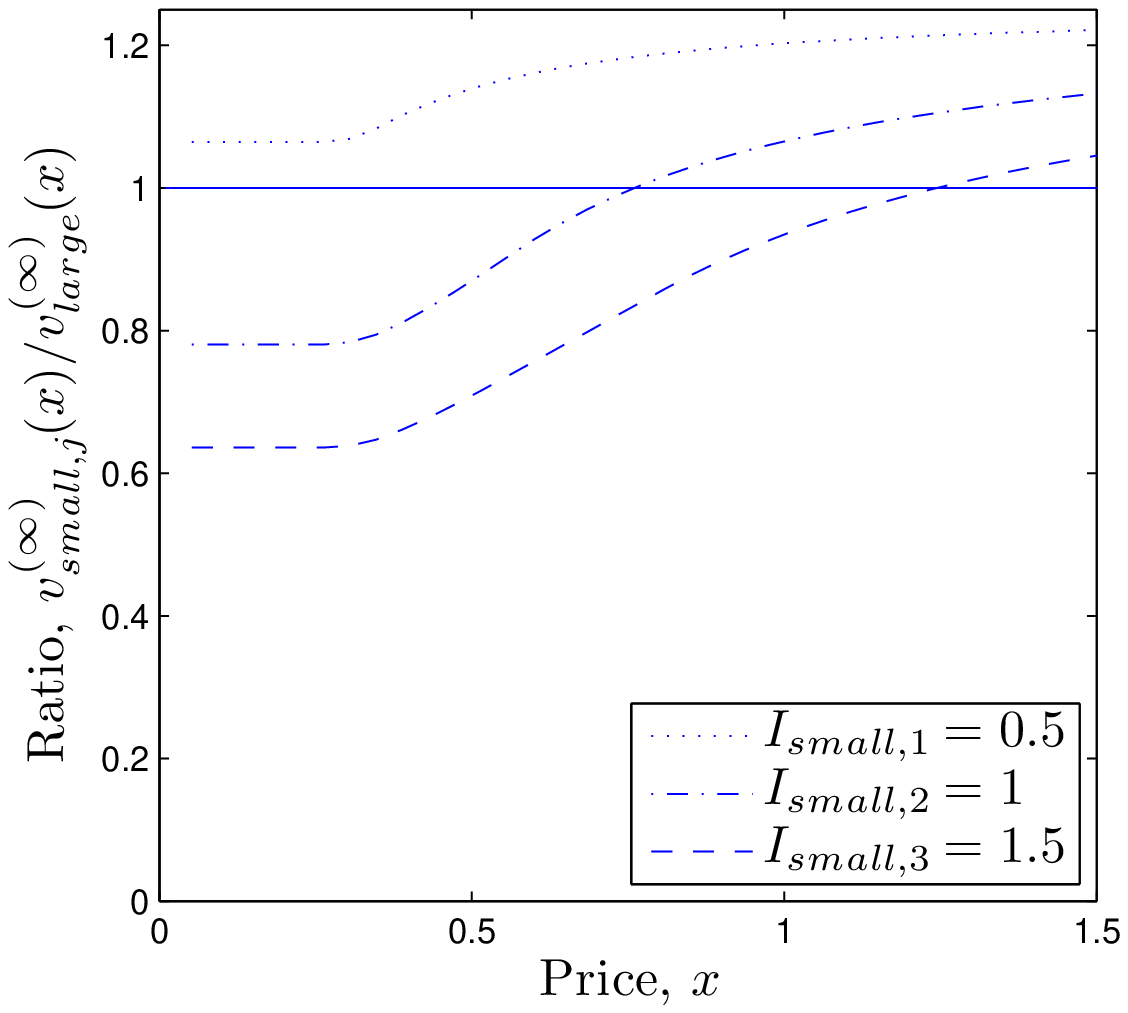}
\end{center}
\caption{\small{(Left) The value of a large-scale investment scenario  ($T_{large}=25$ and $\nu_{small}=5$) together with small-scale analogues ($T_{small}=3$ and $\nu_{small}=0.25$) at different investment costs. The figure verifies the result that the scenario with the shorter lead time has the higher value for large prices, $x$. (Right) Displaying the ratio of the previous value functions $v^{(\infty)}_{small,j}(x)/v^{(\infty)}_{large}(x)$ more clearly reveals that  $v^{(\infty)}_{small,1}(x)>v^{(\infty)}_{large}(x)$ for every $x$. This suggests the existence of a critical investment cost $I_{crit}$ such that the value of a small-scale investment scenario exceeds the traditional large-scale counterpart for any price $x$, as long as $I_{small}\leq I_{crit}$. }}
\label{fig:value_ratio}
\end{figure}

In Figure \ref{fig:value_ratio}, the value $v^{(\infty)}_{large}(x)$ of the large-scale parameter ensemble is displayed alongside $v^{(\infty)}_{small,j}(x)$ with $T_{small}=3$, $\nu_{small}=0.25$ for three different investment costs, $I_{small,j}=0.5,1,1.5$. With their shorter lead time, the values $v^{(\infty)}_{small,j}(x)$ is seen to exceed $v^{(\infty)}_{large}(x)$ for large values of $x$, verifying the analysis above.

\begin{figure}[ht]
\centering
\includegraphics[width=3.3in]{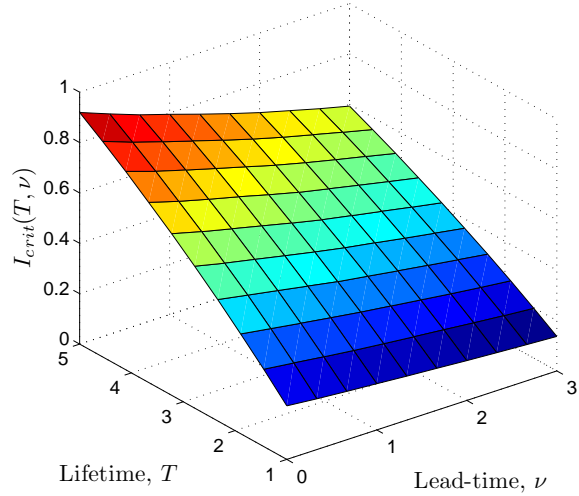}
\caption{\small{Critical investment cost $I_{crit}(T,\nu)$ of a single small-scale investment, with the given $T$ and $\nu$, in order for the value of multiple consecutive such investments to exceed the value of multiple consecutive large-scale investments at any price level. Since the value function, for any choice of $T$ and $\nu$, is decreasing in the investment cost, we have $v^{(\infty)}(I,T,\nu)\geq v^{(\infty)}(I_{large},T_{large},\nu_{large})$, provided that $I<I_{crit}(T,\nu)$. The large-scale investment was characterized by the parameter values: $I_{large}=1$, $T_{large}=25$ years and $\nu_{large}=5$ years.}}
\label{fig:I_max}
\end{figure}

From Theorem \ref{main_theorem} we know that $v^{(\infty)}(x)$ is proportional to $x^{\gamma}$ for small enough values of $x$. This explains the constant appearance of the ratio $v^{(\infty)}_{small,j}(x)/v^{(\infty)}_{large}(x)$ in Figure \ref{fig:value_ratio} for small $x$. The value of the scenario with the lowest investment cost $v^{(\infty)}_{small,1}$ clearly exceeds the value of the `large' scenario for all prices $x$. Indeed, given $T_{large}$, $\nu_{large}$ and $I_{large}$ we can find a critical value $I_{crit}(T_{small},\nu_{small})$, such that the value of the `small' scenario is greater at any price level, as long as $I_{small}<I_{crit}$. This leads to the contour plot in Figure \ref{fig:I_max}, which displays the critical values for different lifetimes, $T_{small}$ and lead times, $\nu_{small}$. In Section \ref{sec:sensitivity},  it has been observed that the value $v^{(\infty)}$ increases with lifetime (see   Figure \ref{fig:lifetime_value})  and decreases with lead time (see Figure \ref{fig:leadtime}). This helps explain  the   trend of $I_{crit}$ with respect to $T$ and $\nu$ in Figure \ref{fig:I_max}. Precisely, in this domain of short lifetimes,  a reduction in lead time yields  a higher critical investment cost, which in turn increases the competitiveness of the small-scale approach.

As an example, with $T_{small}=2.5$ years and $\nu_{small}=0.3$ we can infer from Figure \ref{fig:I_max}  that $I_{crit}=0.5$. That is, despite a difference of a factor 10 in lifetime ($T_{large}=25$ years), the investment cost is required to  differ only by a factor of 2 between the short and  long lifetime scenarios in order for the short-lived one to be preferable.

The closest resemblance of an estimation of $I_{crit}$ using a simplistic NPV argument without any optionality would be to compare the discounted investment costs over an infinite horizon. That is, we can assume that capital is replaced every $T_{large}$ (resp. $T_{small}$) years under the long (resp. short) lifetime scenarios. For simplicity, disregarding features like lead times, operational flexibility, stochastic prices, and optimal multiple stopping, we equate  the discounted costs over an infinite horizon to get
\[\sum_{k=0}^{\infty}e^{-r k T_{large}}I_{large} =\sum_{k=0}^{\infty}e^{-r k T_{small}}I'_{crit}\quad\Leftrightarrow\quad \frac{I'_{crit}}{I_{large}} = \frac{1-e^{-rT_{small}}}{1-e^{-rT_{large}}},\]
where $I'_{crit}$ denotes the critical investment cost of the short-lived scenario under the NPV argument.  Using the same example as above, with $T_{small} = 2.5$ years,  $T_{large} = 25$ years, and  $I_{large}=1$, we  find that $I'_{crit} = 0.24$. Comparing to  the critical investment cost $I_{crit} = 0.50$ above,  the value of the optionality in this particular example permits twice the investment cost that would be suggested by standard net present cost arguments.

Lastly, we note that, in the example with the car engine and the power plant, the cost per kW of capacity of the car engine is almost two orders of magnitude less than that of the power plant \citep{Larminie2003}. Clearly, the critical costs suggested in Figure \ref{fig:I_max} are not nearly as dramatic. This suggests great potential in abandoning the customized, large-scale investments in favor of mass-produced and modular capital.

\section{Conclusions}
\label{extensions}

We have developed a  framework for valuing repeated infrastructure investments  and   comparing traditional capital with long lifetimes and lead times to a mass-produced version with a much shorter lifespan and  lead times. We found that including future investments in present valuations significantly affects the exercise boundary, especially when the individual investment is short-lived. Also, the marginal benefit of increasing lifetime rapidly diminishes when considering multiple investments. Furthermore, this framework allows for a determination of a critical investment cost of capital with lifetimes and lead times that deviate from industry standards. One of examples (see Figure \ref{fig:I_max}) reveals that reducing lifetime of capital from a typical 25 years to 2.5 years need only be accompanied by a decrease of a factor of 2 in investment cost in order to be superior, in overall value terms.

 For our analysis and tractability, we have worked with   a lognormal stochastic cash flow. Working with the lognormal process, we obtain analytical value function for the single stopping problem and analytical results like Theorem 1.  Alternatively, for more realistic representation of commodity price dynamics, other processes, e.g. with  mean reversion and jumps, can be used, though these settings may not be as amenable to mathematical analysis  and the numerical implementation can  be much more challenging and cumbersome. For instance, under the exponential Ornstein-Uhlenbeck  price dynamics, the single stopping problem does not admit a closed-form solution, and the optimal exercise threshold can only be found from an implicit equation (see \cite{LeungLiZhengXOU}). However, under mean reversion, the value of an investment with considerable lead time and lifetime is expected to  generate a cash flow close to the finite long-run   mean. Shorter lived, and more quickly deployed capital would, on the other hand, be more suited to both exploit positive deviations from the mean, and also avoid periods of low prices. In the mean reversion framework, we would therefore anticipate that  the critical investment cost, $I_{crit}$, of the small-scale investment to achieve parity between the small-scale  and large-scale scenarios to  be  higher than those found in Figure \ref{fig:I_max}. This suggests that  a small-scale and modular approach may be attractive  to large infrastructure investments.

There are a number of directions for future research.  First, one can incorporate the firm's aversion to risk and ambiguity associated with the stochastic investment returns, as discussed in \cite{Henderson05} and \cite{jaimungal2011irreversible} in the case of single investment.  Moreover,  additional information on the technological change, industry outlook, and future  investment cost will enhance the decision analysis.  For instance, intuition suggests that  frequent  technological advancement is more easily harnessed with a more rapid turnover since outdated technology can be abandoned without sacrificing investments with a long remaining horizon.  Similarly, regulatory risk is less of an issue with shorter investment windows. This issue is very much in focus in the energy sector today with the looming threat of climate change and related concerns about carbon emissions. For instance, burning cleaner than coal and oil, natural gas is touted as a bridge fuel to cleaner technologies in the future. However, in the current paradigm of large-scale units with long lifetimes, the dramatic increase in natural gas-fired capacity in recent years means a commitment to a fossil fuel for at least another quarter of a century.

\subsection*{Acknowledgements}
The authors are grateful to Klaus Lackner, Garrett van Ryzin and Caner G\"{o}\c{c}men for their useful suggestions and conversations. Tim Leung's research is partially supported by NSF grant DMS-0908295.

\appendix
\section{ Proof of Proposition \ref{prop:positive_reward}}
\label{app:proof_prop}

For every fixed stopping rule  $\vec{\tau}=(\tau_i)_{i= 1}^k\in\mathcal{S}^k$, we define a subsequence $(\hat{\tau}_j)_{j=1}^s$, $s\leq k$, of $(\tau_i)_{i= 1}^k$ by  recording only those stopping times  at which  the reward is non-negative:
\begin{equation*}
\left\{\hat{\tau}_j\right\} = \left\{\tau_i \,|\, \psi(X_{\tau_i}^{0,x})\geq 0 \,;\,i=1,\dots,k\right\}.
\end{equation*}In the case of finite $k$ we can append the subsequence $(\hat{\tau}_j)_{j=1}^s$ with $k-s$ infinite stopping times, $\hat{\tau}_{s+1}=\dots=\hat{\tau}_k=\infty$ to create the  full sequence $\vec{\hat{\tau}}=(\hat{\tau}_j)_{j= 1}^k$. Since the stopping times within  $\vec{\tau}$ are refracted with at least the constant time $T$, the same is true for those in $\vec{\hat{\tau}}$ by construction, and therefore $\vec{\hat{\tau}}\in \mathcal{S}^k$.

 By avoiding those stopping times with a negative reward, the total discounted reward:
\begin{equation*}
g_k(x;\vec{\tau},\psi) := \sum_{i=1}^ke^{-r \tau_i}\psi(X_{\tau_i}^{0,x}),\quad \vec{\tau}\in\mathcal{S}^k,
\end{equation*}
  is dominated by $g_k(x;\vec{\hat{\tau}},\psi)$ in expectation, namely,
\begin{equation}
\label{domination}
\mathbb{E}\left\{g_k(x;\vec{\tau},\psi)\right\}\leq \mathbb{E}\left\{g_k(x;\vec{\hat{\tau}},\psi)\right\}.
\end{equation}
In addition,  when  $\vec{\tau}$ is taken to be $\vec{\hat{\tau}}$, we have the equality $g_k(x;\vec{\hat{\tau}},\psi)=g_k(x;\vec{\hat{\tau}},\psi^+)$, almost surely. That means that maximizing over the stopping rules $\vec{\hat{\tau}}$ for the original problem $v^{(k)}(x)$ will achieve the upper bound (RHS of \eqref{eqnprop1}) with the non-negative reward $\psi^+$. In summary, we obtain
\begin{equation*}v^{(k)}(x)\equiv
\sup_{\substack{\vec{\tau}\in {\cal{S}}^{k}}} \mathbb{E}\left\{g_k(x;\vec{\tau},\psi)\right\}=
\sup_{\substack{\vec{\hat{\tau}}\in {\cal{S}}^{k}}} \mathbb{E}\left\{g_k(x;\vec{\hat{\tau}},\psi^+)\right\}=\sup_{\substack{\vec{\tau}\in {\cal{S}}^{k}}}\mathbb{E}\left\{g_k(x;\vec{{\tau}},\psi^+)\right\}. \quad \square
\end{equation*}

\section{Proof of Theorem \ref{main_theorem}}
\label{app:proof_thm}
\noindent {\bf Proof:} Be the definition of the reward function with a break-even point $x_0$, we know that $\psi(x)<0$ and $\psi '(x)\geq 0$ for  $x<x_0$, and $\psi '(x_0)>0$. This implies that
\begin{equation*}
\Lambda \psi(x) = \gamma\psi(x)-x\psi '(x)<0,\ \ x\leq x_0.
\end{equation*}
Particularly, since $\Lambda \psi(x)$ is convex on $(x_0,\infty)$ and increasing for large $x$, there is exactly one solution, $x_1^*$, to $\Lambda\psi(x_1^*)=0$, and furthermore, $(d/dx)\Lambda\psi(x_1^*)>0$ for $x\geq x_1$. Recall from Lemma \ref{Lemma_1} that
\begin{equation}
\label{v_1}
v^{(1)}(x) = \psi^{(1)}(x\vee x^*_1)\left[1\wedge \left(\frac{x}{x_1^*}\right)^{\gamma}\right],
\end{equation}
where $\psi^{(1)}\equiv \psi$. In addition,
\begin{equation}
\label{psi_2}
\psi^{(2)}(x) = \psi^{(1)}(x) +e^{-r T}\mathbb{E}\left\{v^{(1)}(X_{T}^{0,x})\right\}.
\end{equation}
Since $\Lambda\mathbb{E}\left\{g(X_t)\right\} = \mathbb{E}\left\{\Lambda g(X_t)\right\}$ for any integrable function $g$, we apply (\ref{psi_2}) to get
\begin{eqnarray}
\Lambda\psi^{(2)}(x) &=&\Lambda\psi^{(1)}(x)+ e^{-r T}\mathbb{E}\left\{\Lambda v^{(1)}(X_{T}^{0,x})\right\}\nonumber\\
&=&\Lambda\psi^{(1)}(x)+ e^{-r T}\mathbb{E}\left\{\Lambda \psi^{(1)}(X_{T}^{0,x})\indic{X_t\geq x_1^*}\right\},\label{lambda_reward_2}
\end{eqnarray}
where in the second step we have used the fact that $\Lambda v^{(1)}(x)$ vanishes in the continuation region of $v^{(1)}(x)$, i.e. for $x<x^*_1$. Since $\Lambda \psi^{(1)}(x)$ is assumed convex on $(x_0,\infty)$, the expectation $\mathbb{E}\left\{\Lambda \psi^{(1)}(X_t^{0,x})\indic{X_t\geq x_1^*}\right\}$ is also convex on $(x_0,\infty)$. Being the sum of two convex functions, $\Lambda\psi^{(2)}(x)$ is also convex on $(x_0,\infty)$. Moreover, since $\Lambda \psi^{(1)}(x)\indic{x\geq x_1^*}$ is an increasing function (and strictly positive for $x>x_1^*$), (\ref{lambda_reward_2}) implies that $\Lambda\psi^{(2)}(x)$ is increasing for large enough $x$.

We observe from (\ref{v_1}) and (\ref{psi_2}) that $\psi^{(2)}(x)$ is a continuously differentiable increasing function with $\lim_{x\to 0}\psi^{(2)}(x) =\lim_{x\to 0}\psi^{(1)}(x)<0$. Furthermore, if $\psi^{(1)}(x)$ is bounded by  $f(x) = ax$, for some  $a>0$, then one can show that $v^{(1)}(x)\leq ax$, and therefore
\begin{equation}
\label{bound_psi_2}
\psi^{(2)}(x)\leq ax+e^{-r T}\mathbb{E}\left\{aX_{T}^{0,x}\right\}\leq a\left(1+e^{-(r-\alpha)T}\right)x.
\end{equation}
This ensures the existence of a maximum at $x=x_2^*$ to the function $\psi^{(2)}(x)/x^{\gamma}$, and consequently also the existence of a solution to $\Lambda \psi^{(2)}(x)=0$.  Also,  Proposition \ref{prop:positive_reward} implies that $x_2^*\ge x_0$ as it is never optimal to exercise with negative payoff. The convexity of $\Lambda\psi^{(2)}(x)$ on $(x_0,\infty)$ ensures that there is exactly one such maximum, uniquely defined by
\begin{equation*}
\Lambda\psi^{(2)}(x_2^*) = 0,\ \ {\rm and}\ \ \frac{d}{dx}\Lambda\psi^{(2)}(x) >0,\quad  x \geq x_2^*.
\end{equation*}
Hence, applying Lemma \ref{Lemma_1} again we obtain
\begin{equation*}
v^{(2)}(x) = \psi^{(2)}(x\vee x^*_2)\left[1\wedge \left(\frac{x}{x_2^*}\right)^{\gamma}\right].
\end{equation*}
Also, from the bound on $\psi^{(2)}(x)$ in (\ref{bound_psi_2}) we infer that
$v^{(2)}(x)\leq a\left(1+e^{-(r-\alpha)T}\right)x$.

Repeating the argument above, we can similarly show the existence and uniqueness of a stopping boundary $x_k^*$, for every $k\geq 1$, and also derive an upper bound for each $v^{(k)}$:
\begin{equation}
\label{bound_v}
v^{(k)}(x)\leq ax\left(\sum_{i=0}^{k-1}e^{-(r-\alpha)iT}\right).
\end{equation}

Next, we proceed to prove by induction that $\Lambda\psi^{(k+1)}>\Lambda\psi^{(k)}$, which together with the previous part of the proof would imply that $x^*_{k+1}<x^*_{k}$. As previously remarked, $\Lambda \psi^{(1)}(x)\indic{x\geq x_1^*}$ is positive for $x>x_1^*$. From (\ref{lambda_reward_2}) it follows that
\begin{equation*}
\Lambda\psi^{(2)}(x) -\Lambda\psi^{(1)}(x) = e^{-r T}\mathbb{E}\left\{\Lambda \psi^{(1)}(X_{T}^{0,x})\indic{X_t\geq x_1^*}\right\}>0,
\end{equation*}
Now, assume that $\Lambda\psi^{(k)}(x)>\Lambda\psi^{(k-1)}(x)$. From the definition of $\psi^{(k)}(x)$ in (\ref{intermediate_reward}) we have
\begin{eqnarray}
\Lambda\psi^{(k+1)}(x) - \Lambda\psi^{(k)}(x)&=&e^{-r T}\mathbb{E}\left\{\Lambda v^{(k)}(X_{T}^{0,x}) - \Lambda v^{(k-1)}(X_{T}^{0,x})\right\}\nonumber\\
&=&e^{-r T}\mathbb{E}\left\{\Lambda \psi^{(k)}(X_{T}^{0,x})\indic{x_k^*\leq X_t\leq x_{k-1}^*} \right\}\nonumber\\
&&+e^{-r T}\mathbb{E}\left\{\left(\Lambda \psi^{(k)}(X_{T}^{0,x}) - \Lambda \psi^{(k-1)}(X_{T}^{0,x})\right)\indic{x_{k-1}^*\leq X_t}\right\}\nonumber\\
&>&0.\nonumber
\end{eqnarray}

As for the monotonicity of the sequence $(v^{(k)})_{k\geq 1}$, we first show that $v^{(2)}>v^{(1)}$. Since $v^{(1)}>0$ from Lemma \ref{Lemma_1}, it follows from (\ref{intermediate_reward}) that
\begin{equation}
\psi^{(2)}(x)-\psi^{(1)}(x) = e^{-r T}\mathbb{E}\left\{ v^{(1)}(X_{T}^{0,x})\right\}>0.\label{step1}
\end{equation}
By the fact that  $x_2^*<x_1^*$ and $\psi^{(2)}(x)>\psi^{(1)}(x)$,  the inequality  \[v^{(2)} = \psi^{(2)}(x)>\psi^{(1)}(x)=v^{(1)}\] holds for  $x\geq x^*_1$. Moreover, since $\psi^{(2)}(x)/x^{\gamma}$  is maximized at $x=x^*_2$, we see that
\begin{eqnarray}
v^{(2)}(x)&=&\psi^{(2)}(x)\geq\psi^{(2)}(x_1^*)\left(\frac{x}{x_1^*}\right)^{\gamma}>
\psi^{(1)}(x_1^*)\left(\frac{x}{x_1^*}\right)^{\gamma} = v^{(1)}(x),\ \ x\in(x^*_2,x^*_1).\nonumber
\end{eqnarray}
Similarly, for $x\leq x^*_2$,
\begin{equation}
v^{(2)}(x) = \psi^{(2)}(x_2^*)\left(\frac{x}{x_2^*}\right)^{\gamma}\geq
\psi^{(2)}(x_1^*)\left(\frac{x}{x_1^*}\right)^{\gamma}>
\psi^{(1)}(x^*_1)\left(\frac{x}{x^*_1}\right)^{\gamma}=v^{(1)}(x),\ \ x<x_2^*\,.\label{step2}
\end{equation}
Hence, we have shown that $v^{(2)}>v^{(1)}$. By induction, we obtain
\begin{equation*}
\psi^{(k)}(x) - \psi^{(k-1)}(x)=e^{-r T}\mathbb{E}\left\{v^{(k-1)}(X_{T}^{0,x}) - v^{(k-2)}(X_{T}^{0,x})\right\}>0.
\end{equation*}
With this inequality we can follow the steps from \eqref{step1} to \eqref{step2} to arrive at $v^{(k)}(x)>v^{(k-1)}(x)$. Finally,  the inequality in (\ref{bound_v}) implies that the value function $v^{(k)}$ admits the following bound for any $k$:
\begin{equation*}
v^{(k)}(x)\leq \frac{aM}{1-e^{-(r-\alpha)T}}, \quad x<M. \qquad\ \square
\end{equation*}

\bibliographystyle{elsarticle-harv}
\bibliography{References}




%% \bibitem must have one of the following forms:
%%   \bibitem[Jones et al.(1990)]{key}...
%%   \bibitem[Jones et al.(1990)Jones, Baker, and Williams]{key}...
%%   \bibitem[Jones et al., 1990]{key}...
%%   \bibitem[\protect\citeauthoryear{Jones, Baker, and Williams}{Jones
%%       et al.}{1990}]{key}...
%%   \bibitem[\protect\citeauthoryear{Jones et al.}{1990}]{key}...
%%   \bibitem[\protect\astroncite{Jones et al.}{1990}]{key}...
%%   \bibitem[\protect\citename{Jones et al., }1990]{key}...
%%   \harvarditem[Jones et al.]{Jones, Baker, and Williams}{1990}{key}...
%%


% \end{thebibliography}

\end{document}